\documentclass[epj]{svjour}

\usepackage{latexsym,graphics,amsfonts,amssymb,theorem}

\def\bbbc{{\mathbb C}}

\def\bbbr{{\mathbb R}}

\def\openone{\leavevmode\hbox{\small1\kern-0.355em\normalsize1}}
\def\re{\mbox{\sf Re\,}}
\def\cn{\mbox{cn\,}}
\def\dn{\mbox{dn\,}}
\def\sn{\mbox{sn\,}}
\def\tr{\mbox{tr\,}}
\def\diag{\mbox{diag\,}}
\def\cotan{\mbox{cotan\,}}
\def\arctan{\mbox{arctan\,}}
\def\const{\mbox{const\,}}
\def\rank{\mbox{rank\,}}

\def\bbbe{{\mathbb E}}
\def\Ad{\mbox{Ad\,}}
\newcommand{\hook}{\mathbin{\hbox{$\!\!$
   \vrule height 0.04em width .4em }%
   \hbox{\vrule height .5em width 0.04em}\,}}
\def\openone{\leavevmode\hbox{\small1\kern-0.355em\normalsize1}}
\def\re{\mbox{Re\,}}
\def\tr{\mbox{tr\,}}
\def\im{\mbox{Im\,}}

\begin{document}
\title{Real Hamiltonian forms of Hamiltonian systems}
\author{V. S. Gerdjikov\inst{1,3} \and A. Kyuldjiev\inst{1} \and
G. Marmo\inst{2} \and G. Vilasi\inst{3}%
}                     % Do not remove
%
%\offprints{G. Marmo}          % Insert a name or remove this line
%
\institute{Institute for Nuclear Research and Nuclear Energy, 72
Tzarigradsko chauss\'ee, 1784 Sofia, Bulgaria,\\
\email{gerjikov@inrne.bas.bg, kyuljiev@inrne.bas.bg} \and
Dipartimento di Scienze Fisiche, Universit\`a Federico II di
Napoli and Istituto Nazionale di fisica Nucleare, Sezione di Napoli,
Complesso Universitario di
Monte Sant'Angelo, Via Cintia, 80126 Napoli, Italy,
\email{marmo@na.infn.it} \and
Dipartimento di Fisica "E.R.Caianiello", Universita di Salerno
Istituto Nazionale di fisica Nucleare, Gruppo Collegato di Salerno,
Salerno, Italy, \email{vilasi@sa.infn.it}}
\date{Received: date / Revised version: date}
\abstract{We introduce the notion of a real form of a Hamiltonian
dynamical system in analogy with the notion of real forms for
simple Lie algebras. This is done by restricting the complexified
initial dynamical system to the fixed point set of a given
involution. The resulting subspace is isomorphic (but not
symplectomorphic) to the initial phase space. Thus to each real
Hamiltonian system we are able to associate another nonequivalent
(real) ones. A crucial role in this construction is played by the
assumed analyticity and the invariance of the Hamiltonian under
the involution. We show that if the initial system is Liouville
integrable, then its complexification and its real forms will be
integrable again and this provides a method of finding new
integrable systems starting from known ones. We demonstrate our
construction by finding real forms of dynamics for the Toda chain
and a family of Calogero--Moser models. For these models we also
show that the involution of the complexified phase space induces a
Cartan-like involution of their Lax representations.
\PACS{{02.30.Ik}{Integrable systems}   \and
      {45.20.Jj}{Lagrangian and Hamiltonian mechanics}%
     } % end of PACS codes
} %end of abstract
\maketitle

\section{Introduction}\label{sec:Int}

Recently the so-called complex Toda chain (CTC) was shown to
describe $N $-soliton interactions in the adiabatic approximation
\cite{GUzuEvDi*97,GEI,Arn*99}. The complete integrability of the
CTC is a direct consequence of the integrability of the real
(standard) Toda chain (TC); it was also shown that CTC allows
several dynamical regimes that are qualitatively different from
the one of real Toda chain \cite{GEI}. These results as well as
the hope to understand the algebraic structures lying behind the
integrability of CTC (such as, e.g.\ Lax representation) were the
stimulation for the present work.

We start from a standard (real) Hamiltonian system
$\mathcal{H}\equiv \{\omega ,H,\mathcal{M}\} $ with $n $ degrees
of freedom and Hamiltonian $H$ depending analytically on the
dynamical variables. It is known that such systems can be
complexified and then written as a Hamiltonian system with $2n $
(real) degrees of freedom. Our main aim is to show that to each
compatible involutive automorphism $\widetilde{\mathcal{C}}$ of
the complexified phase space we can relate a real Hamiltonian form
of the initial system with $n $ degrees of freedom. Just like to
each complex Lie algebra one associates several inequivalent real
forms, so to each $\mathcal{H} $ we associate several inequivalent
real forms $\mathcal{H}_\bbbr \equiv \{\omega_\bbbr
,H_\bbbr,\mathcal{M}_\bbbr\} $. Like the initial system
$\mathcal{H} $, the real form is defined on a manifold $\mathcal{M
}_{\bbbr}$ with $n $ real degrees of freedom. Provided
$\widetilde{\mathcal{C}}(H)=H $ the dynamics on the real form will
be well defined and will coincide with the dynamics on $\mathcal{M
}_{\bbbc}$ restricted to $\mathcal{M }_{\bbbr}$. We show that if the
initial system $\mathcal{H} $ is integrable then its real Hamiltonian
forms will also be integrable. We pay special attention to the
connection with integrable systems and the possibility they offer
to define a class of new integrable systems starting from an
initial one. Recently a procedure to obtain new integrable systems
by composing known integrable ones has been elaborated in the
framework of coproducts \cite{Ragnisco}. Here we are not concerned
with this approach.

Examples of indefinite-metric Toda chain (IMTC) have already been studied
by Kodama and Ye \cite{ky}. In particular they note that while the
solutions of the TC model are regular for all $t $, the solutions of the
IMTC model develop singularities for finite values of $t $.  Particular
examples of non-standard (or ``twisted'') real forms of $1+1 $-dimensional
Toda field theories have already been studied by Evans and  Madsen
\cite{Evans} in connection with the problem of positive kinetic energy
terms in the Lagrangian description and with emphasis on conformal WZNW
models.

The approach we follow here is different and more general than the
ones in \cite{ky,Evans}. Its main ideas were reported in \cite{We};
here we elaborate the proofs the details and provide new classes of
examples.

\section{Complexified Hamiltonian Dynamics}\label{sec:CHD}

We start with a real Hamiltonian system with $n$ degrees of
freedom  $\mathcal{H} \equiv \left\{ \mathcal{M}^{(n)} , H  ,
\omega \right\}$ where $\mathcal{M}^{(n)}$ is a $2n$ dimensional
vector space and \begin{equation}\label{w}\omega = \sum_{k=1}^{n}
d p_k \wedge d q_k\end{equation} Let's consider its
complexification:
\begin{equation}\label{eq:CDS}
\mathcal{H}^\bbbc \equiv \left\{ \mathcal{M}^\bbbc , H ^\bbbc ,
\omega^\bbbc \right\}
\end{equation}
where
$\mathcal{M}_\bbbc ^{(2n)}$ can
be viewed as a linear space $\mathcal{M}^{(n)}$ over the field of
complex numbers:
\[ \mathcal{M}^{(2n)}_\bbbc = \mathcal{M}^{(n)}\oplus i
\mathcal{M}^{(n)}.
\]
In other words the dynamical variables $p_k$, $q_k$ in
$\mathcal{M}^{(n)}_\bbbc $ now may take complex values. We assume that
observables $F$, $G$ and the Hamiltonian $H$ are real analytic functions
on $\mathcal{M} $ and can naturally be extended to
$\mathcal{M}^{(2n)}_\bbbc$.

The complexification of the dynamical variables $F$, $G$ and $H$ means
that they become analytic functions of the complex arguments:
\begin{equation}\label{eq:p-q-com}
{p_k}^\bbbc =p_{k,0}+i p_{k,1} , \;\; {q_k}^\bbbc =q_{k,0}+i
q_{k,1} , \;\; k=1,\dots,n_\pm
\end{equation}
and we can write:
\begin{equation}\label{eq:H-com}
H^\bbbc = H(p_k^\bbbc, q_k^\bbbc) = H_0 + i H_1.
\end{equation}
The same goes true also for the complexified 2--form:
\begin{equation}\label{eq:Om-com}
\omega^\bbbc = \sum_{k=1}^{n} d p_k^\bbbc \wedge d q_k^\bbbc =
\omega_0 + i \omega_1
\end{equation}

Note that each of the symplectic forms $\omega _0 $ and $\omega
_1$ are non-degenerate. However the linear combination
$\omega^\bbbc = \omega_0+i\omega_1 $ can be written down in the
form:
\[\omega = \sum_{k=1}^n \left( d p_{k,0}, d p_{k,1},
d q_{k,0}, d q_{k,1} \right)B_0 \left( \begin{array}{c} d
p_{k,0}\\ d p_{k,1}\\ d q_{k,0}\\ d q_{k,1}
\end{array} \right)
\]
where the matrix $B_0 $
\[ B_0 = {1 \over 2} \left(\begin{array}{cccc}
 0 & 0 & 1 & i \\
 0 & 0 & i & -1 \\
 -1 & -i & 0 & 0 \\
 -i & 1 & 0 & 0 \end{array} \right)
 \]
obviously has the property $B_0^2 = 0 $.

\begin{remark}\label{rem:AH}
The kernel of $\omega^\bbbc$ is spanned by the antiholomorphic
vector fields. We could also choose the antiholomorphic
(anti-analytic) functions in the complexification procedure. This
would lead to equivalent results.

\end{remark}

Obviously, $\dim \mathcal{M}^\bbbc =4n $ and therefore
$\mathcal{H}^\bbbc$ may be considered as a real dynamical system
with $2n $ degrees of freedom. To elaborate on this, we start from
the complexified equations of motion:
\begin{eqnarray}\label{eq:eq-mot}
{d p_k^\bbbc \over d t } &=& - {\partial H^\bbbc \over
\partial q_k^\bbbc }, \\ {d q_k^\bbbc \over d t } &=& {\partial
H^\bbbc \over
\partial p_k^\bbbc } \nonumber
\end{eqnarray}
The right hand side of (\ref{eq:eq-mot}) contain the partial
derivatives of both $H_0 $ and $H_1 $. Since we assumed
analiticity, $H_0 $ and $H_1 $ will satisfy the Cauchy-Riemann
equations:
\begin{eqnarray}\label{eq:CR}
&& {\partial H_0 \over \partial q_{k,0}} ={\partial H_1 \over
\partial q_{k,1}}, \qquad
{\partial H_0 \over \partial q_{k,1}} =-{\partial H_1 \over
\partial q_{k,0}}
\end{eqnarray}
which means that the derivatives in the right hand sides of
(\ref{eq:eq-mot}) are equal to:
\begin{eqnarray}\label{eq:}
&& {\partial H^\bbbc \over \partial q_{k}^\bbbc} ={\partial H_0
\over \partial q_{k,0}} - i {\partial H_0 \over \partial q_{k,1}}
={\partial H_1 \over \partial q_{k,1}} + i {\partial H_1 \over
\partial q_{k,0}}.
\end{eqnarray}
 Analogous formulae hold for the derivatives with respect to $p_{k,0} $
 and $p_{k,1} $.
Thus all terms in the right hand sides of (\ref{eq:eq-mot}) can be
expressed through the partial derivatives of $H_0 $ only:
\begin{eqnarray}\label{eq:H_0}
{d p_{k,0} \over d t} = -{\partial H_0 \over \partial q_{k,0}},&
\qquad & {d q_{k,0} \over d t} = {\partial H_0 \over
\partial p_{k,0}}, \nonumber\\ {d p_{k,1} \over d t} = {\partial
H_0 \over \partial q_{k,1}},& \qquad & {d q_{k,1} \over d t} =
-{\partial H_0 \over \partial p_{k,1}}
\end{eqnarray}
Obviously (\ref{eq:H_0}) are standard Hamiltonian equations of
motion for a dynamical system with $2n$ degrees of freedom
corresponding to:
\begin{eqnarray}\label{eq:H_0-om}
  H_0 &=& \re H^\bbbc (p_k^\bbbc, q_k^\bbbc),\qquad\\
 \omega_0 &=& \re \omega^\bbbc = \sum_{k=1}^{n} \left(
d p_{k,0} \wedge d q_{k,0} - d p_{k,1} \wedge d q_{k,1}
\right)\nonumber
\end{eqnarray}
We denote the related real dynamical vector field by $\Gamma_0$.

The system (\ref{eq:eq-mot}) allows a second Hamiltonian formulation with:
\begin{eqnarray}\label{eq:H_1=om}
  H_1 &=& \im H^\bbbc (p_k^\bbbc, q_k^\bbbc), \qquad \\ \omega_1
  &=& \im \omega^\bbbc = \sum_{k=1}^{n} \left( d p_{k,0} \wedge d
q_{k,1} + d p_{k,1} \wedge d q_{k,0} \right)\nonumber
\end{eqnarray}
and also real dynamical vector field $\Gamma_1$. Due to the
analyticity of $H^{\bbbc}$ Cauchy-Riemann equations yield that
these two vector fields actually coincide: $$\Gamma_0 = \Gamma_1\,
.$$ So $\Gamma_0$ is a bi-Hamiltonian vector field and the
corresponding recursion operator is:
\begin{eqnarray}
T&=&\omega_0^{-1}\circ \omega_1 =\\ &-& \frac{\partial}{\partial
p_1}\otimes d p_0 + \frac{\partial}{\partial p_0} \otimes d p_1 -
\frac{\partial}{\partial q_1}\otimes d q_0 +
\frac{\partial}{\partial q_0} \otimes d q_1 \nonumber\\ T^2 &=&
-1.\nonumber
\end{eqnarray}

\section{Complexification and Liouville integrability}\label{sec:LI}

Let us now assume that our initial system is Liouville integrable and
analyze what consequences will have this on the complexified system.

\begin{proposition}\label{pro:LI}
Let the initial system $\mathcal{H} $ have $n$ functionally
independent integrals $I_k$, $k=1,\dots, n$ which are real
analytic (meromorphic) functions of $q_k$ and $p_k$. Let also $I_k
$ be in involution:
\begin{equation}\label{eq:Inv}
\{ I_k, I_j\} \equiv \sum_{s=1}^{n} \left(
{\partial I_k \over \partial p_s} {\partial I_j \over \partial q_s} -
{\partial I_k \over \partial q_s} {\partial I_j \over \partial p_s}
\right) =0,
\end{equation}
Then the complexified system  $\left\{ \mathcal{M}^\bbbc , H_0 , \omega_0
\right\}$ is also Liouville integrable, i.e.\ it has $2n$ functionally
independent integrals $I_{k,0}$, $I_{k,1}$, $k=1,\dots, n$ in involution.
\end{proposition}

\begin{proof}
Obviously after the complexification each of the integrals $I_k $
becomes complex-valued $I_k^\bbbc = I_{k,0} + i I_{k,1} $. Since
$I_k $ is a real analytic function then $I_k^\bbbc $ satisfies
Cauchy-Riemann equations with respect to each of the complexified
variables. Keeping this in mind let us complexify the dynamical
variables in (\ref{eq:Inv}). Separating the real and the
imaginary parts and making use of Cauchy-Riemann equations we get
by direct calculation that:
\begin{eqnarray}\label{eq:I_k-inv}
\{ I_{k,0}, I_{j,0}\}_0 = \{ I_{k,0}, I_{j,1}\}_0 = \{ I_{k,1},
I_{j,1}\}_0 =0,
\end{eqnarray}
where
\begin{eqnarray}\label{eq:PB_0}
\left\{ F, G \right\}_0 &\equiv& {\displaystyle \sum_{s=1}^{n}}
\left( {\partial F \over
\partial p_{s,0}} {\partial G \over \partial q_{s,0}}- {\partial F
\over \partial q_{s,0}} {\partial G \over \partial
p_{s,0}}-\right. \\ && \left. -{\partial F \over
\partial p_{s,1}} {\partial G \over \partial q_{s,1}}+ {\partial F
\over \partial q_{s,1}} {\partial G \over \partial p_{s,1}}
\right) =0,\nonumber
\end{eqnarray}
are the Poisson brackets related to the symplectic form $\omega _0 $
(\ref{eq:H_0-om}).

Quite analogously we can prove that this set of integrals are in
involution also with respect to the Poisson brackets $\{\cdot , \cdot\}_1
$ related to the symplectic form $\omega _1 $. Thus we have proved that
$\left\{ \mathcal{M}^\bbbc , H_0 , \omega_0 \right\}$ has $2n$
integrals in involution.

The next step is to prove that these $2n $ integrals are functionally
independent provided the initial ones $I_k $ are. The independence of $I_k
$ can be expressed by
\begin{equation}\label{eq:I_k-ind}
d I_1\wedge \dots \wedge d I_n \equiv \sum_{i_1<\dots <i_n}^{}
W_{i_1,\dots,i_n} d z_{i_1} \wedge \dots \wedge d z_{i_n} \neq 0,
\end{equation}
where $ W_{i_1,\dots,i_n} $ is the minor of order $n $ of the
$n\times 2n $ matrix $W $ with matrix elements
\begin{equation}\label{eq:W}
W_{jk} = {\partial I_j \over \partial z_k }, \qquad j=1,\dots , n , \quad
k=1,\dots , 2n.
\end{equation}
determined by the columns $1\leq i_i < \dots < i_n \leq 2n $.
Here and below we will denote by $z $ the $2n $-component vector with
components $z_i=q_i $ and $z_{i+n}=p_i $ for $i=1,\dots,n $. In other
words the independence of $I_k$ means that $\rank W=n $.

In the complexified case both the integrals $\mathcal{I}_k\equiv
I_k^\bbbc$  and the dynamical variables $z_j $ become complex-valued.
Their independence can be formulated by:
\begin{eqnarray}\label{eq:2-my-}
 &d\mathcal{I}_1 \wedge \dots \wedge d\mathcal{I}_n \wedge
 d\mathcal{I}_1^* \wedge \dots \wedge d\mathcal{I}_n^*=&
 \\&={\displaystyle \sum_{k=1}^{2n}\sum_{{i_1<\dots <i_k} \atop {j_{k+1} <\dots
<j_{2n-k} }}^{} }W_{i_1,\dots,i_k,j_{k+1},\dots,j_{2n-k}} d
z_{i_1} \wedge \dots \wedge d z_{i_k}\wedge& \nonumber\\& \wedge
\/ d z_{j_{k+1}}^* \wedge \dots \wedge d z_{j_{2n-k}}^* \neq
0,&\nonumber
\end{eqnarray}
where
\begin{eqnarray}\label{eq:3-my}
\mathcal{W} &=& \left( \begin{array}{cc}
 W^{11} & W^{12} \\ W^{21} & W^{22} \end{array} \right), \\
W^{11}_{jk}= {\partial \mathcal{I}_j \over \partial z_k}, &\qquad&
W^{22}_{jk}= {\partial \mathcal{I}_j^* \over \partial z_k^*} =
(W^{11}_{jk})^*, \nonumber \\
W^{12}_{jk}= {\partial \mathcal{I}_j \over \partial z_k^*},
&\quad & W^{21}_{jk}= {\partial\mathcal{I}_j^* \over \partial z_k}
= (W^{12}_{jk})^*. \nonumber
\end{eqnarray}
Therefore we have to prove now that $\rank \mathcal{W}=2n $. But
due to the analyticity of all $I_k $ the matrix elements of
$W^{12}$ and $W^{21}$ vanish and $\mathcal{W} $ becomes a
block-diagonal matrix each block being of rank $n $. The
proposition is proved.
\end{proof}

\begin{remark}\label{rem:AA}
If the transition to action-angle variables is not analytic then
Proposition \ref{pro:LI} is not directly applicable.

\end{remark}

\section{Hamiltonian Reductions and Real Hamiltonian forms}\label{sec:RHF}

In this Section we will show how we associate with each
Hamiltonian system $\mathcal{H}$ a {\em family of RHF}. We will do
this by using a special type of reductions.

There are several methods to reduce Hamiltonian systems to systems
with lower number of degrees of freedom. One of the best known is
the reduction by using integrals of motion \cite{Marmo}. Indeed,
if we constrain our dynamical system by fixing the values of $p$
integrals of motion $I_j(q_k,p_k) =\const $, $j=1,\dots,p $, then such
constraint is compatible with the dynamics and leads to reduced
system with $n-p$ degrees of freedom. Here, we will follow somewhat
different route by restricting the dynamics to ${\mathcal M}_\bbbr$ which
is the fixed point set of a Cartan-like involutive automorphism
$\widetilde{\mathcal{C}}$, defined below.

The approach we will follow in this paragraph is inspired by the basic
idea of construction of real forms for simple Lie algebras \cite{Lie}. A
basic tool in our construction is the automorphism
$\widetilde{\mathcal{C}}$ , which plays the r\^ole of a ``complex
conjugation operator''.

Let us introduce involutive canonical automorphism
$\mathcal{C}$ (i.e.\ ${\mathcal C}^2=1\,$, ${\mathcal C}^*\omega =
\omega$) on the phase space $\mathcal{M}^{(n)}$ and on its
dual\footnote{In general we should have different notations of
$\mathcal{C}$ in these spaces \cite{Marmo}. However, since our
phase space is a vector space with some abuse of notations we will
use the same letter for both realizations. } by:
\begin{eqnarray}\label{eq:J} && \mathcal{C} \left( \{ F, G\}
\right)= \left\{ \mathcal{C}(F), \mathcal{C} (G) \right\}, \qquad
\mathcal{C}^2=\openone,
\end{eqnarray}
where $F$, $G \in \mathcal{F}(\mathcal{M}^{(n)})$ are real
analytic functions on $\mathcal{M}^{(n)}$. The involution acts on
them by:
\begin{eqnarray}\label{eq:1'}
&&\mathcal{C} (F(p_1,\dots ,p_n,q_1,\dots ,q_n))\nonumber\\
&&=F((\mathcal{C}(p_1),\dots ,\mathcal{C}(p_n),\mathcal{C}(q_1),\dots ,
\mathcal{C}(q_n))).
\end{eqnarray}

In terms of vector fields $X, Y \in T\mathcal{M}$ and the lifted
involution $T\mathcal{C} : T\mathcal{M }\rightarrow T\mathcal{M }$
we have:
\begin{eqnarray}\label{eq:2}
{\omega (T\mathcal{C}(X),T\mathcal{C}(Y)) = \mathcal{C}(\omega
(X,Y))},\quad (T\mathcal{C})^2=\openone
\end{eqnarray}
where $\omega$ is the symplectic form corresponding to the Poisson
brackets in (\ref{eq:J}).

Since $\mathcal{C}$ has eigenvalues $1$ and $-1$, it naturally
splits $\mathcal{M}^{(n)}$ into two subspaces $\mathcal{M}^{(n)}=
\mathcal{M}^{(n_+)}_+\oplus \mathcal{M}^{(n_-)}_-$ such that
\begin{eqnarray}\label{eq:Mpm}
\mathcal{C}{\bf X}= {\bf X} \quad &\mbox{for}& \quad {\bf X} \in
\mathcal{M}^{(n_+)}_+ , \nonumber \\ \mathcal{C}{\bf Y}= -{\bf Y}
\quad &\mbox{for}& \quad {\bf Y}\in \mathcal{M}^{(n_-)}_-
\end{eqnarray}
where $n_\pm = \dim \mathcal{M}^{(n_\pm)}_\pm $, $ n=n_+ + n_-$. We also
assume that the starting Hamiltonian $H$ is invariant with respect to
$\mathcal{C}$:
\begin{equation}\label{eq:CH-in}
\mathcal{C} (H) = H.
\end{equation}

Equation (\ref{eq:J}) guarantees that each of the subspaces
$\mathcal{M}^{(n_+)}_+ $, $\mathcal{M}^{(n_-)}_- $
is a symplectic subspace of $\mathcal{M}^{(2n)}_\bbbc$.
On $\mathcal{M}^{(n_+)}_+ $ and $\mathcal{M}^{(n_-)}_- $ we have
\[ \omega_+ =\sum_{k=1}^{n_+} d p_k^+ \wedge d q_k^+ . \qquad
 \omega_- =\sum_{k=1}^{n_-} d p_k^- \wedge d q_k^-
\]
where $p_k^+$, $q_k^+$ (resp. $p_k^-$, $q_k^-$) are basis in
$\mathcal{M}^{(n_+)}_+ $ (resp. $\mathcal{M}^{(n_-)}_- $). With
respect to the automorphism $\mathcal{C}$ they satisfy:
\begin{equation}\label{eq:Cp-q}
\mathcal{C}(p_k^\pm) =\pm p_k^\pm, \qquad \mathcal{C}(q_k^\pm)
=\pm q_k^\pm
\end{equation}
for all $k=1,\dots,n_\pm$.

In $\mathcal{M}^{(n)}_\bbbc $ along with $\mathcal{C}$ we can
introduce also the complex conjugation $*$. In this construction
obviously $\mathcal{C}$ commutes with $*$ and their composition
$\widetilde{\mathcal{C}} = \mathcal{C}\circ * = *\circ
\mathcal{C}$ is again an involutive automorphism on
$\mathcal{M}^{(2n)}_\bbbc$.

\begin{corollary}\label{cor:1}
Let $I_j $, $j=1,\dots , n $ be $n $ integrals of motion in involution of
our initial system $\mathcal{H} $, depending analytically on $q_k $ and
$p_k $.  Let us denote their extensions to $\mathcal{M}^\bbbc $ by
$\mathcal{I}_j =I_{j,0}+iI_{j,1} $.  Then both $\mathcal{I}_j $ and
\begin{equation}\label{eq:I_j^*}
\widetilde{\mathcal{C}}(\mathcal{I}_j) = \mathcal{I}_j^*
\end{equation}
are also integrals of motion in involution for the complexified system
$\mathcal{H}^\bbbc $.
\end{corollary}

Using eq. (\ref{eq:J}) it is easy to check that the dynamics
generated by a Hamiltonian which is invariant with respect to
$\mathcal{C} $ (\ref{eq:CH-in}) will have the subspaces
$\mathcal{F}_+ $ and $\mathcal{F}_- $ of $\mathcal{C}$-invariant
and $\mathcal{C}$-antiinvariant functions as invariant subspaces.
Moreover $\mathcal{F}_+ $ turns out to be also an invariant
subalgebra.

The real form $\mathcal{M}^{(n)}_\bbbr $ of the phase space is the
symplectic (with respect to $\omega_0$) subspace of
$\mathcal{M}^{(2n)}_\bbbc $ invariant with respect to
$\widetilde{\mathcal{C}}$:
\begin{equation}\label{eq:4}
\mathcal{M}^{(n)}_\bbbr = \mathcal{M}^{(n_+)}_+ \oplus i
\mathcal{M}^{(n_-)}_-
\end{equation}
Indeed any element of $\mathcal{M}_{\bbbr}^{(n)}$ can be represented as:
\begin{equation}\label{eq:Mr}
\vec{Z}=\vec{X}+i\vec{Y} \in \mathcal{M }_{\bbbr}^{(n)}.
\end{equation}
where $\vec X$ and $\vec Y$ are real-valued elements of
$\mathcal{M}^{(n_+)}_+ $ and $\mathcal{M}^{(n_-)}_-$ respectively. The
reality condition means that
\begin{equation}\label{eq:Zr}
\tilde{\mathcal{C}}(\vec{Z})\equiv \mathcal{C}(\vec{Z}^*) =
\mathcal{C}(\vec{X}-i\vec{Y}) =\vec{X}+i\vec{Y} =\vec{Z}
\end{equation}
where we have made use of Eq. (\ref{eq:Mpm}).

Here and below by $F_\bbbr$ and $G_\bbbr\in
\mathcal{F}(\mathcal{M}^{(n)}_\bbbr) $ we denote the restriction
of the observables $F,\,G$ by restricting their arguments to
$\mathcal{M}_\bbbr$. Then $F_\bbbr$ and $G_\bbbr$ satisfy the
analog of Eq. (\ref{eq:1'}) with $\mathcal{C}$ replaced by
$\widetilde {\mathcal{C}}$.  Due to Eq. (\ref{eq:J}) their Poisson
bracket $\{F_\bbbr,G_\bbbr\} \in
\mathcal{F}(\mathcal{M}^{(n)}_\bbbr)$ too, i.e.\
$\mathcal{F}(\mathcal{M}^{(n)}_\bbbr)$ becomes a Poisson
subalgebra. If we choose the Hamiltonian $H_\bbbr\in
\mathcal{F}(\mathcal{M}^{(n)}_\bbbr)$ then
\begin{equation}\label{eq:H-inv}
\widetilde {\mathcal{C}}(H_\bbbr) = H_\bbbr.
\end{equation}
The evolution associated with $H_\bbbr$ of the dynamical variable
$F_\bbbr$:
\begin{eqnarray}\label{eq:Jcom}
{dF_\bbbr \over d t }= \{ H_\bbbr,F_\bbbr\},
\end{eqnarray}
defines a dynamics on $\mathcal{M }_{\bbbr}^{(n)} $. Rewriting
(\ref{eq:Jcom}) into its equivalent form:
\begin{equation}\label{eq:7'}
\omega (X_{H_{\bbbr}},\cdot ) = dH_\bbbr \cdot
\end{equation}
and making use of (\ref{eq:2}) we see that the vector field $X_{H_\bbbr} $
must also satisfy
\begin{equation}\label{eq:7''}
\widetilde{C} (X_{H_\bbbr}) = X_{\widetilde{C}(H_\bbbr)}=
X_{H_\bbbr}
\end{equation}
on $\mathcal{M}^{(n)}_\bbbr$. The symplectic form restricted on
$\mathcal{M }_{\bbbr}^{(n)} $ also becomes real and equals
\begin{equation}\label{ome_R}
\omega_\bbbr =\sum_{k=1}^{n_+} d p_{k,0}^+ \wedge d q_{k,0}^+ -
\sum_{k=1}^{n_-} d p_{k,1}^- \wedge d q_{k,1}^-
\end{equation}
where $p_{k,0}^+$, $ q_{k,0}^+$, $k=1,\dots ,n_+$ and $p_{k,1}^-$,
$ q_{k,1}^-$, $k=1,\dots ,n_-$ are the basic elements in
$\mathcal{M}_\bbbr^{(n)}$.

If $\mathcal{M}_{\bbbc}^{(2n)} $ is endowed with Hamiltonian which is
``real'' with respect to $\widetilde{\mathcal{C}}$ and whose vector
field $X_H$ satisfies (\ref{eq:7''}) then the restriction of the
dynamics on $\{\mathcal{M }_{\bbbr}^{(n)}$, $\omega _{\bbbr}$,
$H_\bbbr \}$ is well defined and coincides with the dynamics on
$\{\mathcal{M }_{\bbbc}^{(2n)}, \omega_0, H_0\}$ restricted to
$\mathcal{M }_{\bbbr}^{(n)} $.

The complexified equations of motion take the form:
\begin{eqnarray}\label{eq:8eqs}
{d p_{k,0}^+ \over d t} = -{\partial H_0 \over \partial
q_{k,0}^+},& \qquad & {d p_{k,1}^+ \over d t} = {\partial H_0
\over
\partial q_{k,1}^+}, \nonumber\\
{d p_{k,1}^- \over d t} = {\partial H_0 \over \partial
q_{k,1}^-},& \qquad & {d p_{k,0}^- \over d t} = -{\partial H_0
\over \partial q_{k,0}^-}, \nonumber\\ {d q_{k,0}^+ \over d t} =
{\partial H_0 \over \partial p_{k,0}^+},& \qquad & {d q_{k,1}^+
\over d t} = -{\partial H_0 \over \partial p_{k,1}^+}, \\
{d q_{k,1}^- \over d t} = -{\partial H_0 \over
\partial p_{k,1}^-},& \qquad & {d q_{k,0}^- \over d t} =
{\partial H_0 \over
\partial p_{k,0}^-} \nonumber
\end{eqnarray}

We shall show that the dynamics on $\{\mathcal{M }_{\bbbc}^{(2n)}
$, $\omega_0$, $H_0\}$ reduces naturally to ${\mathcal M}_\bbbr$
which is the fixed point set of $\widetilde{\mathcal{C}}$.

\begin{proposition}\label{pro:2}
Let the Hamiltonian system $\mathcal{H}^\bbbc $ and the involution
$\tilde{\mathcal{C}} $ be as above. Then the second class Dirac
constraints
\begin{equation}\label{eq:M_red}
p_{k,1}^+=0, \qquad q_{k,1}^+=0, \qquad p_{k,0}^-=0, \qquad
q_{k,0}^-=0
\end{equation}
on the subspace $\mathcal{M}_\bbbr $ invariant with respect to
$\tilde{\mathcal{C}} $ are compatible with the dynamics of the
complexified Hamiltonian system $\mathcal{H}^\bbbc $. The reduced
equations of motion take the form:
\begin{eqnarray}\label{eq:4eqs}
&& {d p_{k,0}^+ \over d t} = -{\partial H_\bbbr \over
\partial q_{k,0}^+}, \nonumber\\ && {d p_{k,1}^- \over d t} =
{\partial H_\bbbr \over \partial q_{k,1}^-}, \\ && {d q_{k,0}^+
\over d t} = {\partial H_\bbbr \over \partial p_{k,0}^+},
\nonumber\\ && {d q_{k,1}^- \over d t} = -{\partial H_\bbbr \over
\partial p_{k,1}^-} \nonumber
\end{eqnarray}
where $H_\bbbr = H_0|_{\mathcal{M}_\bbbr}$.
\end{proposition}

\begin{proof}

Since $H$ is real analytic function, then
\[
H_0 (\dots, p_{k,1}^+, q_{k,1}^+,\dots)= H_0 (\dots, -p_{k,1}^+,
-q_{k,1}^+,\dots)
\]
is an even function of $p_{k,1}^+$ and $ q_{k,1}^+$. Besides from
the condition $C(H)=H$ it follows, that
\[
H_0 (\dots, p_{k,0}^-, q_{k,0}^-,\dots)= H_0 (\dots, -p_{k,0}^-,
-q_{k,0}^-,\dots)
\]
and $H_0$ is an even function also of $p_{k,0}^-$ and $q_{k,0}^-$.
Obviously the first derivatives of $H_0$ with respect to these
variables are odd functions and therefore:
\begin{equation}\label{eq:9.1H}
\left. {\partial H_0 \over \partial q_{k,1}^+}
\right|_{\mathcal{M}_\bbbr} = \left. {\partial H_0 \over \partial
q_{k,0}^-} \right|_{\mathcal{M}_\bbbr} = \left. {\partial H_0
\over \partial p_{k,1}^+} \right|_{\mathcal{M}_\bbbr} = \left.
{\partial H_0 \over \partial p_{k,0}^-}
\right|_{\mathcal{M}_\bbbr} =0 .
\end{equation}
The real Hamiltonian form is determined by:
\[ \mathcal{H}_\bbbr \equiv \left\{ \mathcal{M}_\bbbr ,
H _\bbbr(p,q), \omega_\bbbr \right\},
\]
\[
\mathcal{M}_\bbbr = \mathcal{M}_+ \oplus i\mathcal{M}_-, \qquad
H_\bbbr(p,q)= H_0(p_k,q_k)|_{\mathcal{M}_\bbbr},
\]
\[
\omega_\bbbr= \omega_0|_{\mathcal{M}_\bbbr} = \sum_{k=1}^{n_+} d
p_{k,0}^+ \wedge d q_{k,0}^+ - \sum_{k=1}^{n_-}d\tilde{p}_{k,1}^-
\wedge d\tilde{q}_{k,1}^-
\]
where $d\tilde{p}_{k,1}^- = -i d p_{k,1}^-$, $ d\tilde{q}_{k,1}^-
=-id q_{k,1}^-$. The proposition is proved.
\end{proof}

Thus we have proved that the equations (\ref{eq:8eqs}) can be
consistently restricted to $\mathcal{M}_\bbbr$ and give rise to a
well defined dynamical system with $n$ degrees of freedom
$\mathcal{H}_\bbbr \equiv \left\{ \mathcal{M}_\bbbr, \omega_\bbbr,
H_\bbbr \right\}$ which we call a real Hamiltonian form (RHF) of the
initial Hamiltonian system $\mathcal{H}\equiv \left\{ \mathcal{M}, \omega,
H \right\}$.

Let us now consider the set of observables $F_\bbbr $, $G_\bbbr $
related to our RHF $\mathcal{H}_\bbbr $. They can be obtained from
the observables of the complexified system $\mathcal{H}^\bbbc $ by
restricting their variables to $\mathcal{M}_\bbbr $.  We remind
that we are selecting the class of observables which depends
analytically on the dynamical variables. This means, that after
restricting on $\mathcal{M}_\bbbr $ they satisfy
\begin{equation}\label{eq:9.1F}
\left. {\partial F_0 \over \partial q_{k,1}^+}
\right|_{\mathcal{M}_\bbbr} = \left. {\partial F_0 \over \partial
q_{k,0}^-} \right|_{\mathcal{M}_\bbbr} = \left. {\partial F_0
\over \partial p_{k,1}^+} \right|_{\mathcal{M}_\bbbr} = \left.
{\partial F_0 \over \partial p_{k,0}^-}
\right|_{\mathcal{M}_\bbbr} =0
\end{equation}
and analogous relations for the partial derivatives of $G $.
If we now calculate the Poisson brackets (\ref{eq:PB_0}) between two
observables $F_\bbbr $ and $G_\bbbr $ we easily find that due to eqs.
(\ref{eq:9.1F}) they will simplify to
\begin{eqnarray}\label{eq:PB_R}
  \{ F_\bbbr, G_\bbbr\}_0 &\equiv & \sum_{s=1}^{n_+} \left(
{\partial F_\bbbr \over \partial p_{s,0}^+} {\partial G_\bbbr
\over \partial q_{s,0}^+}- {\partial F_\bbbr \over \partial
q_{s,0}^+} {\partial G_\bbbr \over \partial p_{s,0}^+} \right)\\
&-&\! \sum_{s=n_++1}^{n} \!\left( {\partial F_\bbbr \over
\partial p_{s,1}^-} {\partial G_\bbbr \over \partial q_{s,1}^-}-
{\partial F_\bbbr \over
\partial q_{s,1}^-} {\partial G_\bbbr \over \partial p_{s,1}^-}
\nonumber\right)
\end{eqnarray}
Obviously the Poisson brackets (\ref{eq:PB_R}) correspond to the
symplectic form $\omega _\bbbr $.

\begin{proposition}\label{pro:LI_R}
The RHF $\mathcal{H}_\bbbr $ corresponding to a Liouville integrable
Hamiltonian system $\mathcal{H} $ is Liouville integrable.
\end{proposition}

\begin{proof}
We start with a $\mathcal{H} $ which has $n $ integrals in involution
$I_k$ depending analytically on the dynamical variables. The
complexification provides us with $2n $ integrals of motion $\mathcal{I}_k
$ and $\mathcal{I}_k^* $ which are also in involution. Let us now restrict
ourselves to $\mathcal{M}_\bbbr $. To this end we use the involution
$\widetilde{\mathcal{C}} $. It is easy to check that $n $ of the integrals
are preserved:
\begin{equation}\label{eq:**0}
\mathcal{I}_{k,\bbbr}^+ = {1  \over 2 } \left. \left(
\mathcal{I}_k + \widetilde{\mathcal{C}} (\mathcal{I}_k)\right)
\right|_{\mathcal{M}_\bbbr}
\end{equation}
and become invariant with respect to $\widetilde{\mathcal{C}} $ while the
other $n $ integrals vanish:
\begin{equation}\label{eq:**1}
\mathcal{I}_{k,\bbbr}^- = {1  \over 2 } \left. \left( \mathcal{I}_k -
\widetilde{\mathcal{C}} (\mathcal{I}_k)\right)
\right|_{\mathcal{M}_\bbbr} =0,
\end{equation}
The fact that the integrals $\mathcal{I}_{k,\bbbr}^+ $ are in involution
with respect to the Poisson brackets $\{ \cdot , \cdot \}_\bbbr $ follows
from (\ref{eq:PB_R}). The proposition is proved.

\end{proof}

Those $I_j$ which have definite $\mathcal
C$-parity will produce real first integrals $I_{j,\bbbr}$ for the
real form dynamics  and those with minus $\mathcal C$-parity will produce
purely imaginary integrals.

\begin{remark}\label{rem:1}
We proved that after restricting $\mathcal{H}^\bbbc $ to
$\mathcal{M}_\bbbr $ the integrals of motion satisfy (\ref{eq:**1}). One
may ask whether the inverse statement also holds true. Namely, assume that
we restrict $\mathcal{H}^\bbbc$ by using simultaneously all $n $
constraints in (\ref{eq:**1}) together with their symplectic conjugate
ones. Since $\mathcal{I}_{k,\bbbr}^- $ are all independent on
$\mathcal{M}_\bbbc $ then such a procedure would lead to a dynamical
system with $n $ degrees of freedom. Will this new system coincide with
$\mathcal{H}_\bbbr $? We believe that the answer to this question is
positive, though we do not yet have a rigorous proof of this fact.
\end{remark}

If we restrict our dynamical variables on $\mathcal{M}_\bbbr $ then only
the integrals invariant with respect to $\widetilde{\mathcal{C}} $
survive.

Let us assume now that we have the complexified system $\mathcal{H}^\bbbc
$ restricted by (\ref{eq:**1}). Let us now pick a point $m\in
\mathcal{M}_\bbbr $ which will correspond to the initial condition of our
dynamical equations. Obviously the evolution can not take this point out
of  $\mathcal{M}_\bbbr $. Indeed, if we assume the opposite we easily
find that the corresponding Hamiltonian vector field is not invariant with
respect to $\widetilde{\mathcal{C}} $. This means that the Hamiltonian $H
$ is not invariant with respect to $\mathcal{C} $ which is a
contradiction.

Following the same line we can prove that if the initial point belongs to
$\mathcal{M}\backslash \mathcal{M}_\bbbr $ then the evolution can not take
it out of $\mathcal{M}\backslash \mathcal{M}_\bbbr $.

As a result the set of constraints (\ref{eq:**1}) split
$\mathcal{M}_\bbbc $ into two orthogonal invariant subspaces.

For some special choices of $H $ it may happen that the dynamics of
$\mathcal{H} $ and $\mathcal{H}_\bbbr $ coincide.

Despite the apparent simplicity of the transition to the real form
dynamics it is essentially different from the initial one---the
real form dynamical vector field satisfying $\Gamma_{\bbbr} \hook
\omega_{\bbbr} = -d H_{\bbbr}$ is generally not even a (locally)
Hamiltonian vector field for the initial $\omega$. As a result we
may expect that the new and the initial dynamical vector fields
will coincide only for very few cases. Conditions for this are
settled in the following

\begin{proposition}\label{pro:3}
$\Gamma_{\bbbr} = \Gamma$ iff

i) the Hamiltonian is separable in a sum of two parts depending on
as follows: $H=H_+(p_k^+,q_k^+) + H_-(p_j^-,q_j^-)$ where $q_k^+$,
$p_k^+$ (resp. $q_j^-$, $p_j^-$) are elements of $\mathcal{M}_+$
(resp. $\mathcal{M}_- $), {\em and}

ii) $H_-(ip_j^-,iq_j^-)=-H_-(p_j^-,q_j^-) $.

Also, condition i) is equivalent both to the local Hamiltonianity
of\/ $\Gamma$ with respect to\/ $\omega_{\bbbr}$ and to the
compatibility of\/ $\Gamma$ with\/ $\omega_{\bbbr}$ i.e.\ ${\mathcal
L}_\Gamma\, \omega_{\bbbr}=0\,$.
\end{proposition}
\begin{proof}
The dynamical vector field satisfying $\Gamma_{\bbbr} \hook
\omega_{\bbbr} = -d H_{\bbbr}$ will be
\begin{eqnarray}\label{eq:Gama_R}
\Gamma_{\bbbr}&=&\sum_{k=1}^{n_+} \left({\partial H_{\bbbr} \over \partial
p_k^+}{\partial \over \partial q_k^+} -{\partial H_{\bbbr} \over \partial
q_k^+}{\partial \over \partial p_k^+}   \right) \nonumber\\
&+& \sum_{j=1}^{n_-} \left( {\partial H_{\bbbr} \over
\partial q_j^-}{\partial \over \partial p_j^-} - {\partial H_{\bbbr}
\over \partial p_j^-}{\partial \over \partial q_j^-}\right)
\end{eqnarray}
and we have to satisfy:
\begin{eqnarray}\label{eq:G=G}
&& {\partial H \over \partial q_k^+}={\partial H_{\bbbr} \over
\partial q_k^+}  \qquad  {\partial H \over \partial p_k^+}
={\partial H_{\bbbr} \over \partial p_k^+}\nonumber\\
&& {\partial H \over \partial q_j^-}=-{{\partial H_{\bbbr}} \over
\partial q_j^-}  \qquad  {\partial H \over \partial p_j^-}=
-{\partial H_{\bbbr} \over \partial p_j^-}.
\end{eqnarray}

Comparing the equations from the first and second rows of (\ref{eq:G=G})
we obtain that all mixed derivatives of $H $ should vanish:
\begin{eqnarray}\label{d2H}
{\partial^2 H \over \partial q_k^+\,\partial q_j^-}&=&0, \qquad
{\partial^2 H \over \partial p_k^+\,\partial p_j^-}=0, \nonumber\\
{\partial^2 H \over \partial q_k^+\,\partial p_j^-}&=&0, \qquad
{\partial^2 H \over \partial p_k^+\,\partial q_j^-}=0
\end{eqnarray}
which is exactly the separability  condition  {\em i)} of the
Hamiltonian. Now {\em ii)} part of the Proposition follows
trivially from the way we construct $H_{\bbbr}$.

Due to the closedness of $\omega$ (and $\omega_{\bbbr}$) we have:
\begin{eqnarray}\label{H-ty}
&&{\cal L}_\Gamma \omega_{\bbbr} = d(\Gamma\hook
\omega_{\bbbr}) \\
&&\; = 2 \sum_{j,k} \left( {\partial^2 H \over \partial q_k^+\,
\partial p_j^-}\, d q_k^+ \wedge d p_j^-  + {\partial^2 H \over \partial
q_j^-\, \partial p_k^+}\, d q_j^- \wedge d p_k^+\right)\nonumber
\end{eqnarray}
and vanishing of all mixed derivatives in eq.(\ref{d2H}) is the
condition for the required compatibility and/or local
Hamiltonianity. Since the separability of $H$ is equivalent to the
separability of $H_{\bbbr}$ we have similarly compatibility of\/
$\Gamma_{\bbbr}$ with\/ $\omega$ and local Hamiltonianity of
$\Gamma_{\bbbr}$ with respect to $\omega$.

\end{proof}

When the conditions of the Proposition \ref{pro:3} are fulfilled our
procedure will give us a bi-Hamiltonian description of the initial
dynamics with a recursion operator $$T=\omega^{-1}\circ \omega_{\bbbr}
=\openone_+ - \openone_-$$ where $\openone_{\pm}$ are the identity tensors
on $\mathcal{M}^{(n_\pm)}_{\pm}$.  So, whatever the outcome we gain either
new dynamics or a bi-Hamil\-tonianity of the old one.

For instance, if we have a collection of uncoupled harmonic
oscillators $H = \sum_i (p_i^2 + q_i^2)$ and a compatible
involution: $\mathcal{C}(p_i)=\epsilon_i p_i $, ${\mathcal
C}(q_i)=\epsilon_i q_i$ with $\epsilon_i=\pm 1$ then we will have
$H_{\bbbr} = \sum_i \epsilon_i(p_i^2 + q_i^2)$ and $\Gamma_{\bbbr} =
\Gamma$\/ i.e.\ coinciding initial and real form dynamics. The
same will be also true for $U(ip_j^-,iq_j^-)= -U(p_j^-,q_j^-)$.

At present, we do not have a receipt how to construct and classify
all involutions of $\mathcal{M} $ consistent with a given Hamiltonian.
However in many important cases it is possible to provide a number of such
non-trivial involutions.

\section{RHF of Completely Integrable systems}\label{sec:CIS}

Here we understand the notion of completely integrable systems in
a broader sense \cite{LanMarVil}. We will say that a dynamical
system with $n $ degrees of freedom is completely integrable if we
find $2n $ functions $\{I_k, \phi _k\} $, $k=1,\dots, n $ such
that $I_k $ are integrals of motion, i.e.:
\begin{equation}\label{eq:L_Ik}
L_{\Gamma } I_k =0, \qquad k=1,\dots, n,
\end{equation}
and $\phi _k $ are nilpotent with respect to the vector field $L_{\Gamma }
$ of order $2 $:
\begin{equation}\label{eq:L_phik}
L_{\Gamma }L_{\Gamma } \phi _k =0, \qquad k=1,\dots, n.
\end{equation}

If there is one or more functions of the $\phi$'s which are
constants of the motion, the system is said to be superintegrable
\cite{BeppeLecce}. We also assume that $\{I_k, \phi _k\} $
introduce local coordinates on our phase space.

Such set of functions $\{I_k, \phi _k\} $ generalize the standard
notion of action-angle variables (AAV) \cite{LanMarVil}. While the
standard AAV can be introduced only for compact motion on a torus,
our variables $\{I_k, \phi _k\} $ can be derived also for
non-compact dynamics. This is important to our purposes,
especially when we analyze how the transition from one RHF to
another changes the character of the motion from compact to a
non-compact one and vice versa, see Section 8 below.   Therefore
from now on, with some abuse of language we will say that $\{I_k,
\phi _k\} $ are our AAV.

The complexification renders all $I_k^\pm$ and
$\phi_k^\pm$ complex: $I_k^{\bbbc,\pm} = I_{k,0}^\pm +i
I_{k,1}^\pm$ and $\phi_k^{\bbbc,\pm} = \phi_{k,0}^\pm
+i\phi_{k,1}^\pm$ and the automorphism $\widetilde{\mathcal{C}}$
has the form:
\begin{equation}\label{eq:Ct-I-f}
\widetilde{\mathcal{C}} (d I_k^{\bbbc,\pm})=\pm (d
I_k^{\bbbc,\pm})^*,\qquad \widetilde{\mathcal{C}} (d
\phi_k^{\bbbc,\pm})=\pm (d \phi_k^{\bbbc,\pm})^*
\end{equation}
The automorphism $\widetilde{\mathcal{C}}$ obviously satisfies the
condition (\ref{eq:7''}). In order to satisfy also
(\ref{eq:CH-in}) we need to assume that $H$ is an even function of
all $I_k^-\in \mathcal{M}_-$. Then the restriction on $\mathcal{M}_\bbbr$
according to (\ref{eq:Mr}) and (\ref{eq:Ct-I-f}) means
restricting all $I_k^+, \phi_k^+ \in \mathcal{M}_+$ to be real, while
all $I_k^-, \phi_k^- \in \mathcal{M}_-$ become purely imaginary.
Then:
\begin{eqnarray}\label{eq:H_R}
H_\bbbr &=& H(I_{1,0}^+, \dots , I_{n_+,0}^+, i I_{1,1}^-, \dots ,
iI_{n_-,1}^-), \nonumber \\
\omega_\bbbr &=& \sum_{k=1}^{n_+} d I_{k,0}^+\wedge d \phi_{k,0}^+
- \sum_{k=1}^{n_-} d I_{k,1}^-\wedge d \phi_{k,1}^-
\end{eqnarray}
and obviously we have again a completely integrable Hamiltonian
system.

Till the rest of this section we also assume that the Hamiltonian
is separable, i.e.
\begin{equation}\label{eq:46'}
H=\sum_{k=1}^{n_+} h_k^+(I_k^+) +\sum_{k=1}^{n_-} h_k^-(I_k^-).
\end{equation}
Obviously the condition (\ref{eq:CH-in}) requires that $h_k^-(I_k^-)$ must
be even functions of $I_k^-$.

Another important approach to completely
integrable systems \cite{Vilasi} is based on the notion of the
recursion operator -- a \mbox{(1,1)} tensor field with vanishing
Nijenhuis torsion \cite{LMV}. This tensor field reflects the
possibility to introduce a second symplectic structure $\omega_1$
on $\mathcal{M}$ through:
\begin{equation}\label{eq:T-d}
 \omega_1(X,Y) = {1\over 2} \left( \omega(TX,Y) + \omega(X,TY)
 \right ).
\end{equation}
In terms of $I_k^\pm$, $\phi_k^\pm$ the tensor field $T$ and
$\omega_1$ can be expressed by:
\begin{eqnarray}\label{eq:T-ome}
&& T = \sum_{k=1}^{n_+} T_k^+ + \sum_{k=1}^{n_-} T_k^- , \qquad
 \omega_1 = \sum_{k=1}^{n_+} \omega_{1,k}^+ + \sum_{k=1}^{n_-}
 \omega_{1,k}^-, \nonumber \\
&& T_k^\pm = \zeta_k^\pm (I_k^\pm) \left( d I_k^\pm \otimes
{\partial \over \partial I_k^\pm} + d\phi_k^\pm \otimes
 {\partial \over \partial \phi_k^\pm} \right) , \nonumber\\
 && \omega_{1,k}^\pm = \zeta_k^\pm (I_k^\pm) d I_k^\pm \wedge
 d\phi_k^\pm
\end{eqnarray}
where $\zeta_k^\pm (I_k^\pm)$ are some functions of $I_k^\pm$; we assume
that they are real analytic functions of their variables.

In order that $\omega_1$ also satisfy eq. (\ref{eq:2}) it is enough that
$\mathcal{C}(T)=T$. In particular this means that $\zeta_k^-$ must be even
functions of $I_k^-$. If this is so then we can repeat our construction
also for $\omega_1$; i.e., we can complexify it and then restrict it onto
$\mathcal{M}_\bbbr$ with the result:
\begin{eqnarray}\label{eq:T-omer}
T_\bbbr &=& \sum_{k=1}^{n_+} T_{\bbbr,k}^+ + \sum_{k=1}^{n_-}
T_{\bbbr,k}^- , \\
T_{\bbbr,k}^+ &=& \zeta_k^+ (I_k^+) \left( d I_k^+\otimes {\partial \over
\partial I_k^+} + d\phi_k^+\otimes {\partial \over \partial \phi_k^+}
 \right) , \nonumber\\
T_{\bbbr,k}^- &=& \zeta_k^- (iI_k^-) \left( d I_k^-\otimes {\partial
\over \partial I_k^-} + d\phi_k^-\otimes {\partial \over \partial
 \phi_k^-} \right) ,  \nonumber\\
\omega_{1,\bbbr} &=& \sum_{k=1}^{n_+} \zeta _k^+(I_k^+)
dI_k^+\wedge d\phi _k^+ - \sum_{k=1}^{n_-} \zeta _k^-(iI_k^-)
dI_k^-\wedge d\phi _k^-.\nonumber
\end{eqnarray}

Thus by construction, the restriction to other real forms
preserves the Nijenhuis property of $T$ and the condition for
double degenerate and nowhere constant eigenvalues. Also it is
easy to check that it is preserved by the dynamical flow:
$\mathcal{L}_{\Gamma_\bbbr}T_{\bbbr}=0$. Due to the fact that
existence of recursion operators is equivalent \cite{dFMSV} to
integrability at least in the non-resonant case, this line of
argumentation gives us another instrument to treat the
integrability of real forms.

The separability of $H$ (\ref{eq:46'}) looks rather restrictive condition.
However, all integrable systems obtained by reducing a soliton
equation (like, e.g. the nonlinear Schr\"odinger equation) on its
$N$-soliton sector are separable. Finite-dimensional systems allowing Lax
pairs like Toda chains, Calogero-Moser systems etc. also possess this
property, see eq. (\ref{eq:I_k}) below.

\section{Examples}\label{sec:Exa}

We illustrate our construction by several paradigmatic examples which
include several types of Toda chain models and Calogero-Moser
models.

\begin{example}
Toda chain related to the $sl(n,\bbbc) $ algebra. We consider
simultaneously the conformal and the affine case by introducing the
parameter $c_0 $ which takes two values: $c_0=0 $ (conformal case) and
$c_0=1 $ (affine case).
\begin{eqnarray}\label{exa:TC}
H_{\rm TC} &=& \sum_{k=1}^{n} {p_k^2 \over 2} + \sum_{k=1}^{n-1}
{e}^{q_{k+1}-q_{k}} + c_0 {e}^{q_{1}-q_{n}},
\nonumber \\
 \omega &=& \sum_{k=1}^n d p_k\wedge d q_k.
\end{eqnarray}
We choose the involution as:
\begin{equation}\label{eq:ex3.1}
\mathcal{C}(p_{k}) = - p_{\bar{k}}, \qquad \mathcal{C}(q_{k}) = -
q_{\bar{k}}.
\end{equation}
where $\bar{k}=n+1-k$.
We  also introduce new symplectic coordinates adapted to the action of
$\mathcal{C}$:
\begin{eqnarray}\label{CA}
p_{k}^+ = {p_k - p_{\bar{k}} \over \sqrt{2}}, &\quad& q_{k}^+ =
{q_k-q_{\bar{k}} \over \sqrt{2}}, \\ \widetilde p_{k}^- = {p_k +
p_{\bar{k}} \over \sqrt{2}}, &\quad& \widetilde q_{k}^- = {q_k
+q_{\bar{k}} \over \sqrt{2}} , \qquad k=1,\dots , r=[n/2]\nonumber
\end{eqnarray}
where $n=2r +r_0$ with $r_0=0$ or $1$; for $r_0=1$ we have to add
also: $\widetilde p = p_{r+1}$, $\widetilde q = q_{r+1} $. These
coordinates are such that
\begin{eqnarray}\label{eq:CAa}
\mathcal{C} (p_{k}^\pm) &=& \pm p_{k}^\pm, \qquad
\mathcal{C} (\widetilde{q}_{k}^\pm) =\pm \widetilde {q}_{k}^\pm, \\
{\mathcal C} (\widetilde{p}) &=& -\widetilde p, \qquad  \mathcal{C}
(\widetilde{q}) = -\widetilde q \nonumber
\end{eqnarray}
As a result we obtain the following real forms of the TC model:
i)~for $n=2r+1 $:
\begin{eqnarray}\label{eq:ex3.4a}
H_{\rm TC1}&=& {1\over 2}\sum_{k=1}^{r} ((p_{k}^+)^{2} -
(p_{k}^-)^{2}) - {1\over 2}(p_{r+1}^-)^2  +\nonumber\\&+&
2\sum_{k=1}^{r-1} {e}^{(q_{k+1}^+ - q_{k}^+)/\sqrt{2}} \cos
{q_{k+1}^- - q_{k}^- \over \sqrt{2}} \nonumber \\&+&
2{e}^{-q_{r}^+/\sqrt{2}} \cos
\left(q_{r+1}^- - {q_{r}^- \over \sqrt{2}}\right) , \\
\omega_\bbbr &=& \sum _{k=1}^r d p_k^+ \wedge d q_k^+ - \sum
_{k=1}^{r+1} d p_k^- \wedge d q_k^-.\nonumber
\end{eqnarray}

ii)~for $n=2r $:
\begin{eqnarray}\label{eq:ex3.4b}
  H_{\rm TC2}&=& {1\over 2} \sum_{k=1}^{r} ((p_{k}^+)^{2} -
(p_{k}^-)^{2}) + {e}^{-\sqrt{2}q_{r}^+} +\\&+& 2\sum_{k=1}^{r-1}
{e}^{(q_{k+1}^+ - q_{k}^+)/\sqrt{2}} \cos {q_{k+1}^- - q_{k}^-
\over \sqrt{2}}, \nonumber \\ \omega_\bbbr &=& \sum _{k=1}^r d
p_k^+ \wedge d q_k^+ - \sum _{k=1}^{r} d p_k^- \wedge d
q_k^-.\nonumber
\end{eqnarray}

These models are generalizations of the well known Toda chain
models associated to the classical Lie algebras \cite{MiOlPer};
indeed if we put $q_{k}^-\equiv 0 $ and $p_{k}^-\equiv 0 $ we find
that (\ref{eq:ex3.4a}) goes into the ${\bf B}_r $ TC while
(\ref{eq:ex3.4b}) provides the ${\bf C}_r $ TC.
\end{example}

\begin{example}\label{exa:CMS}
 The real Hamiltonian forms for the Calogero--Moser systems (CMS) \cite{CMS}.
The CMS corresponding to the root systems of $A$, $B$, $C$, $D$
and $BC$-series \cite{Perelomov} are defined by Hamiltonians of
the type:
\begin{equation}\label{eq:H-CMS}
H_{\rm CMS}= 1/2 \sum_{j=1}^n p_j^2 + U(q)
\end{equation}
with:
\[
U=\left\{ \begin{array}{ll} g^2 V^n_ -,& \mbox{   for } \qquad A_{n-1} \\
g^2 (V^n_- + V^n_+) +g^2_1 V^n_1 +g^2_2 V^n_2 ,& \mbox{   for } \qquad
BC_{n}.\end{array} \right.
\]
Here
\begin{eqnarray}
V^n_{\pm}= \sum_{j< k \le n} v(q_j\pm q_k), \qquad V^n_1&=&
\nonumber \sum_{j\le n} v(q_j), \\
V^n_2&=& \sum_{j\le n} v(2q_j),
\end{eqnarray}
and the function $v(x)$ is one of the following:
\begin{equation}\label{eq:ex2_v}
v(q) = \frac{1}{q^2},\quad \frac{1}{\sin^2 q},\quad
\frac{1}{\sinh^2 q},\quad \frac{1}{q^2} +\omega q^2,\quad \wp(q)
\end{equation}
where $\wp(q) $ is the Weierstrass function.

The $BC_n $ case contains in itself the CMS for the algebras $B_{n}$,
$C_{n}$ and $D_{n}$; they are obtained by putting $g_2=0$,  $g_1=0$ and
$g_1=g_2=0$ respectively.

All these models are invariant under the involution
(\ref{eq:ex3.1}). For convenience we again use the coordinates
(\ref{eq:CAa}) together with the following complex coordinates
\[
z_j = {1 \over \sqrt{2} } \left( q_{k}^+ + i q_{k}^- \right), \qquad
z_j^* = {1 \over \sqrt{2} } \left( q_{k}^+ - i q_{k}^- \right),
\]
$k=1,\dots , r. $

In order to obtain the Hamiltonian of the RHF of the CMS we have to
express $H_{\rm CMS} $  (\ref{eq:H-CMS}) in the new coordinates and
assume that $\widetilde p_{k}^-$, $\widetilde q_{k}^-$, $\widetilde p$
and $\widetilde q$ are purely imaginary and then calculate its real
part.  We shall denote the imaginary parts of these variables by the same
letter but without tilde and so all tilde-less variables will be real. As
a result:

\begin{eqnarray*}
\re \sum_{j=1}^n {p_j^2 \over 2} &=& {1 \over 2} \sum_{k=1}^{r}
{(p_{k}^+)}^{2}-
{1 \over 2} \sum_{k=1}^{r} {(p_{k}^-)}^{2} - {1 \over 2} r_0 p^2, \\
  \re V^n_{1} &=& 2\re \sum_{k=1}^{r} v(z_j) - r_0 v(q), \\
\re V^n_{2} &=& 2\re \sum_{k=1}^{r} v(2z_j) - r_0 v(2q), \\  \re
V^n_{-} &=& \nonumber 2\re \sum_{j<k\le r} \left[ v(z_j-z_k)+
v(z_j + z_k^*)\right]  +\\&&+ \sum_{k=1}^{r} v(\sqrt{2} q_{k}^+) +
2r_0 \re \sum_{j\le r} v(z_j - i q),\\  \re V^n_{+} &=&
 2\re\! \sum_{j<k\le r}\left[
v(z_j+z_k) + v(z_j-z_k^*) \right] + \\&& -\sum_{k=1}^{r}
v(\sqrt{2} q_{k}^-)+ 2r_0\re\sum_{j\le r} v(z_j+i q).
\end{eqnarray*}
To illustrate the form of the new Calogero-Moser potentials we
note that:
\begin{eqnarray}
\re {1 \over \sin^2 (x+i y)} &=& {(\sin x \cosh y)^2 - (\cos x
\sinh y)^2 \over ((\sin x)^2 + (\sinh y)^2)^2}\, ,\nonumber\\ \re
{1 \over \sinh^2 (x+i y)}&=& {(\sinh x \cos y)^2 - (\cosh x \sin
y)^2 \over ((\sinh x )^2 + (\sin y)^2)^2}. \nonumber
\end{eqnarray}
Together with
\begin{equation}\label{omegarf}
\omega_{\bbbr} = \sum_{k=1}^{r}\left( d p_{k}^+ \wedge d q_{k}^+
- d p_{k}^- \wedge d q_{k}^- \right)- r_0 d p \wedge d q
\end{equation}
they define the real form dynamics which is obviously not confined
to the standard Calogero-Moser models.

\end{example}

More explicitly:

for $n=2r+1$ and $\mathfrak{g}\simeq A_{n-1}$:
\begin{eqnarray*}
  H_{\rm CM,\bbbr} &=& {1\over 2} \sum_{k=1}^r \left[ (p_k^+)^2
-(p_k^-)^2 \right] - {1\over 2} (p_{r+1}^-)^2 +\\&+& 2g^2
\sum_{i<j}^r \re \left[ v(z_i -z_j) + v(z_i +z_j^*)\right] +\\&+&
g^2 \sum_{j=1}^r v(\sqrt{2}q_j^+) + 2g^2 \sum_{j=1}^r \re v(z_i -i
q_{r+1}^-).
\end{eqnarray*}

for $n=2r$ and $\mathfrak{g}\simeq A_{n-1}$:
\begin{eqnarray*}
H_{\rm CM,\bbbr} &=& {1\over 2} \sum_{k=1}^r \left[ (p_k^+)^2
-(p_k^-)^2 \right] + g^2 \sum_{j=1}^r v(\sqrt{2}q_j^+)+\\
&+& 2g^2 \sum_{i<j}^r \re \left[ v(z_i -z_j) + v(z_i
+z_j^*)\right] .
\end{eqnarray*}

for $n=2r+1$ and $\mathfrak{g}\simeq BC_{n}$:
\begin{eqnarray*}
  H_{\rm CM,\bbbr}& &= {1\over 2} \sum_{k=1}^r \left[ (p_k^+)^2
-(p_k^-)^2 \right] - {1\over 2} (p_{r+1}^-)^2 +\\&+& g^2
\sum_{j=1}^r \left[ v(\sqrt{2}q_j^+) - v(\sqrt{2}q_j^-)\right]+ \\
&+& 2g^2 \sum_{i<j}^r \re \left[ v(z_i -z_j)\right. +\\&&+ \left.
v(z_i +z_j^*) + v(z_i +z_j) + v(z_i -z_j^*)\right]+ \\ &+& 2g^2
\sum_{j=1}^r \re \left[ v(z_i -i q_{r+1}^-) + v(z_i+i q_{r+1}^-)
\right] -\\&-& g_S^2 v(q_{r+1}^-) - g_L^2 v(2q_{r+1}^-)+\\&+&
2g_S^2 \sum_{j=1}^r \re v(z_j)+ 2g_L^2 \sum_{j=1}^r \re v(2z_j) .
\end{eqnarray*}

for $n=2r$ and $\mathfrak{g}\simeq BC_{n}$:
\begin{eqnarray*}
H_{\rm CM,\bbbr} &&= {1\over 2} \sum_{k=1}^r \left[ (p_k^+)^2
-(p_k^-)^2 \right] +\\
&+& g^2 \sum_{j=1}^r \left[ v(\sqrt{2}q_j^+)
- v(\sqrt{2}q_j^-) \right] +\\
&+& 2g^2 \sum_{i<j}^r \re \left[ v(z_i -z_j) \right. +\\&&+ \left.
v(z_i +z_j^*) + v(z_i +z_j) + v(z_i -z_j^*)\right]+
\\&+& 2g_S^2 \sum_{j=1}^r \re v(z_j)+ 2g_L^2 \sum_{j=1}^r \re
v(2z_j) .
\end{eqnarray*}
Note that: $v(i y) = - v(y)$.

\begin{example}
$\;$ In the case of involution $\,\mathcal{C}_2 = S_{e_1-e_2}$:
\[ \mathcal{C}_2(q_1) = q_2, \;\; \mathcal{C}_2(q_2) = q_1,
\;\; \mathcal{C}_2(p_1) = p_2, \;\; \mathcal{C}_2(p_2) = p_1,
\]
we obtain for $\mathfrak{g}\simeq BC_{n}$:
\begin{eqnarray*}
  H_{\rm CM,\bbbr} &=& {1\over 2} \left[ (p_1^+)^2 - (p_1^-)^2 +
\sum_{k=3}^n (p_k^+)^2 \right] +\\
&+& g_S^2 \left[ 2\re v(z_1) +
\sum_{j=3}^n \re v(q_j) \right]+\\
&+& g^2 \left\{ 2\sum_{j=3}^n \re \left[ v(z_1- q_j^+ ) + v(z_1 +
q_j^+)\right]\right.-\\
&&-  v(\sqrt{2} q_1^-) + v(\sqrt{2} q_1^+) +\\
&&+\left.  \sum_{3\leq
k<j}^n \left[ v(q_k-q_j) + v(q_k+q_j) \right] \right\}+\\
&+& g_L^2 \left[ 2\re v(2z_1) + \sum_{j=3}^n \re v(2q_j) \right]
\end{eqnarray*}
\end{example}

\begin{example}
$\;$ In the case of involution $\mathcal{C}_3 = -\openone$:
\[ \mathcal{C}_2(q_k) = -q_k, \;\; \mathcal{C}_2(p_k) = -p_k,
\]
\[ q_k^- = q_k, \;\; p_k^- = p_k, \;\; \dim \mathcal{M}_- =
2n. \]

We obtain for Toda chains with the algebra $A_{n-1}\simeq sl(n)$:
\[
H_{\rm TC} = {1\over 2} \sum_{k=1}^n p_k^2 + {m^2 \over \beta^2}
\sum_{k=1}^{n-1} {e}^{\beta (\alpha_k, \vec{q})} + c_0 {e}^{\beta
(\alpha_0, \vec{q})},
\]
where $c_0=0$ stands for conformal TC and $c_0=1$ for affine TC.
Now all $p_k$ and $q_k$ become purely imaginary. We choose also
$\beta = i\beta_0$. As a result we obtain Toda chain with purely
imaginary interaction constant.

Similarly, for CMS we obtain:

A) If $v(q) = 1/q^2$ or $v(q) = 1/q^2 +\omega^2 q^2$ then we
obtain:
\[
H_{\rm CM,\bbbr} = - H, \qquad \mbox{which is equivalent to}
\qquad t \leftrightarrow -t.
\]

B) If $v_2(q) = a^2/ \sin^2(aq)$ and $v_3(q) = a^2/ \sinh^2(aq)$
then the real form dynamics is obtained by the exchange:
\[ v_2(q) \leftrightarrow - v_3(q), \qquad t \leftrightarrow -t.
\]

C) If $v_4(q) = a^2\wp(aq|\omega_1,\omega_2)$ then the real form
dynamics is obtained by the exchange:
\[ v_4(q) \leftrightarrow - v_4'(q), \qquad t \leftrightarrow -t.
\]
where $v_4'(q) = -a^2\wp(aq|i\omega_1,i\omega_2)$.

\end{example}
\begin{remark}\label{rem:Calo}
Several of the RHF of CMS have been obtained earlier by Calogero
\cite{Calo} using the so-called `duplication procedure'. Our results above
show that each `duplication' is related to a Cartan-like automorphism of
the relevant Lie algebra. This fact can be used to classify all
inequivalent RHF of CMS.

\end{remark}

\begin{example}\label{exa:Grav}$\;$
{An example from general relativity: transition from $SO(3)$ to
$SO(2,1)$ symmetry.}

Recall that the geodesic Schwarzschild flow, corresponding to a metric $
g=g_{ij}dx_{i}dx_{j}$ on $\mathcal{M}$, can be viewed as the flow,
generated by the Hamiltonian field $X_{H}$ with $
H=\frac{1}{4}g^{ij}p_{i}p_{j}$ ($p_{i}$'s are conjugated to $x_{i}$'s
coordinates on $T^{\ast }\mathcal{M}$).

For example, the geodesic flow for the Schwarzschild metric
\begin{eqnarray}
g_{1}=&\epsilon _{1}&\left(\frac{rc^{2}-2MG}{rc^{2}} d
t^{2}-\frac{rc^{2}}{ rc^{2}-2MG}\right) d r^{2}+\nonumber\\ &+&
\epsilon _{2}r^{2}\left( d\vartheta ^{2}+\sin ^{2}\vartheta\,
d\varphi ^{2}\right) ,
\end{eqnarray}
where $\epsilon _{1}=\pm 1$, $\epsilon _{2}=\pm 1$, is described by the
(completely integrable) Hamiltonian
\begin{eqnarray}
\mathcal{H}_{1}=\frac{1}{4}&&\left[ \epsilon _{1}\left(
\frac{rc^{2}-2MG}{rc^{2}}p_{t}^{2}-\frac{rc^{2}-2MG}
{rc^{2}}p_{r}^{2}\right)\right.+\nonumber\\&&\left.
+\frac{\epsilon _{2}}{r^{2}}\left( p_{\vartheta
}^{2}+\frac{1}{\sin ^{2}\vartheta } p_{\varphi }^{2}\right)
\right] . \label{gf1}
\end{eqnarray}

The complexification and real projection procedure leads to the
following Hamiltonian \begin{eqnarray}
\mathcal{H}_{2}=\frac{1}{4}&&\left[ \epsilon _{1}\left(
\frac{rc^{2}-2MG}{
rc^{2}}p_{t}^{2}-\frac{rc^{2}-2MG}{rc^{2}}p_{r}^{2}\right)\right.\nonumber
-\\&&\left. -\frac{\epsilon _{2}}{r^{2}}\left( p_{\vartheta
}^{2}+\frac{1}{\sinh ^{2}\vartheta } p_{\varphi }^{2}\right)
\right] \end{eqnarray} which (is still completely integrable and)
is associated with the metric
\begin{eqnarray}
g_{2}=\epsilon _{1}&&\left( \frac{rc^{2}-2MG}{rc^{2}}d
t^{2}-\frac{rc^{2}}{ rc^{2}-2MG}\right) d
r^{2}-\nonumber\\&&-\epsilon _{2}r^{2}\left( d\vartheta ^{2}+\sinh
^{2}\vartheta \,d\varphi ^{2}\right) ,
\end{eqnarray}
solution of vacuum Einstein field equations \cite{KKK99,SVV01}.

The above notation for coordinates might be misleading because, in
$g_{1}$ and $\mathcal{H}_{1}$, $r\in \left]0 ,\infty \right[
,\vartheta \in \left] -\pi,\pi \right[ ,\varphi \in \left[ 0,2\pi
\right[ $ denote spherical coordinates, while, in $g_{2}$ and
$\mathcal{H}_{2}$, $r\in \left] 0,\infty \right[ ,\vartheta \in
R,\varphi \in \left[ 0,2\pi \right[ $ denote pseudo-spherical
coordinates.

The $2$-dimensional surfaces endowed with metric $g_{2}$ restricted by $
\left( r,t=\const\right) $ may be identified with one of the sheets of the
two-sheeted space-like hyperboloid. They are also known as \textit{
pseudo-spheres}.

The pseudo-sphere is a surface with constant negative Gaussian curvature $
\mathcal{R}=-1/r^{2}$. It can be globally embedded in a $3$-dimensional
Minkowskian space. Let $y_{1},y_{2},y_{3}$ denote the coordinates in the
Minkowskian space, where the separation from the origin is given by
$y^{2}=-y_{1}^{2}+y_{2}^{2}+y_{3}^{2}$. These coordinates are connected to
the pseudo-spherical coordinates $\left( r,\vartheta ,\varphi \right) $
by:
\[
y_{1}=r\cosh \vartheta ,\qquad y_{2}=r\sinh \vartheta \cos \varphi
,\qquad  y_{3}=r\sinh \vartheta \sin \varphi.
\]
The equation $ y^{2}=-r^{2}$, \textit{i.e.} the locus of points
equidistant from the origin, specifies a hyperboloid of two sheets
intersecting the $y_{1}$ axis at the points $\pm r$ called \textit{poles}
in analogy with the sphere. Either sheet (say the upper sheet) models an
infinite spacelike surface without a boundary; hence, the Minkowski metric
becomes positive definite (Riemannian) upon it. This surface has
constant Gaussian curvature ($ \mathcal{R}=-1/r^{2}$), and it is
the only simply connected surface with this property. Other
embeddings of the pseudo-sphere in the $3$-dimensional Euclidean
space are also available, for example it can be regarded as the
$2$ -dimensional surface generated by the tractrix, but they are
not global.

It is worth noting that metrics $g_{1}$ and $g_{2}$ are $so\left(
3\right)$-invariant and $so\left( 2,1\right)$-invariant\footnote{
In the pseudo-spherical coordinates, the $so(2,1)$ Lie algebra $
[X_{1},X_{2}]=X_{3},\quad \lbrack X_{2},X_{3}]=-X_{1},\quad
\lbrack X_{3},X_{1}]=-X_{2},$, is spanned by $X_{1}=\sin \varphi
\partial _{\vartheta }+\cos \varphi \coth \vartheta \partial
_{\varphi }$, $ X_{2}=-\cos \varphi
\partial _{\vartheta }+\sin \varphi \coth \vartheta
\partial _{\varphi }$, $X_{3}=\partial _{\varphi }$ \cite{VV01}.},
respectively, and that geodesic flows, corresponding to the model
Ricci-flat metrics, associated with $3$-dimensional Killing algebras, are
integrable. To see that one may observe that, for instance,
Hamiltonian $\mathcal{H}_{2}$ possesses five independent first
integrals
\[
\mathcal{H}_2,\quad p_{\tau },\quad p_{\vartheta }^{2}+\frac{1}{\sinh
^{2}\vartheta }p_{\varphi }^{2},\quad \mathcal{I}_{1},\quad \mathcal{I}
_{2}\quad
\]
where $\mathcal{I}_{1}$ and $\mathcal{I}_{2}$ are generators of a
noncommutative two-dimensional subalgebra of $so\left( 2,1\right) $, for
example
\begin{eqnarray*}
  \mathcal{I}_{1} &=&\left[ \left( 1+\sqrt{2}\right) \cos
\varphi +\sin \varphi \right] p_{\vartheta }+\\&&+\left[
1+\sqrt{2}+\coth \vartheta \left( \cos \varphi -\left(
1+\sqrt{2}\right) \sin \varphi \right) \right] p_{\varphi }
\\
\mathcal{I}_{2} &=&\sqrt{2}\left[ \cos \varphi +\sin \varphi
\right] p_{\vartheta }+\\&&+\left[ 2+\sqrt{2}\coth \vartheta
\left( \cos \varphi -\sin \varphi \right) \right] p_{\varphi }.
\end{eqnarray*}
The $5$ first integrals span a rank $3$ Lie algebra $\mathcal{A}$, and
since
\[
\rank\mathcal{A}+\dim \mathcal{A}=\dim T^{\ast }\mathcal{M}\quad
\mbox{and} \quad \rank\mathcal{A}<\dim \mathcal{A}
\]
the system is noncommutatively integrable in the sense \cite{MF78,SV00}.

Geodesic flows corresponding to the metrics invariant for a
$3$-dimensional Lie algebra are discussed in \cite{SVV01}. Note
also that the above proposition is no more valid for the geodesic
flow of Ricci-flat metrics with nonextendable two-dimensional
Killing algebras.

\end{example}

\section{Lax operators}

It is well known that a number of completely integrable systems
admit Lax representation. By it we mean the existence of two
matrices $L(\vec{p} , \vec{q} )$ and $M(\vec{p}, \vec{q} ) $
depending explicitly on the dynamical variables and such that the
Lax equation
\begin{equation}\label{eq:Lax}
{dL \over d t} = [L,M]
\end{equation}
is equivalent to the corresponding equations of motion of
$\mathcal{H} $. Let us also remind that from (\ref{eq:Lax}) there
follows immediately that the eigenvalues $\zeta _s $ of $L $ are
integrals of the motion in involution. One can use also as such
\begin{equation}\label{eq:I_k}
I_k = \tr L^k(\vec{p},\vec{q}) = \sum_{s=1}^{n} \zeta _{s}^{k},
\qquad k=1,\dots, n,
\end{equation}
and as a rule the Hamiltonian $H $ is a linear combination of
$I_k$'s; for example, for the Toda chain and for the CMS $H= 2\,
I_2 $.

If the matrices $L $ and $M $ are analytic (meromorphic) functions
of $p_k $ and $q_k $ then obviously these properties will hold
true also for the integrals $I_k $. The analytic properties of
$\zeta _s $ as functions of $p_k $ and $q_k $ require
additional considerations. Indeed, $\zeta _s $ are roots of the
corresponding characteristic equation which is of order $n$ so the
mapping from $p_k $ and $q_k $ to $\zeta _s $ may have essential
singularities.

The Lax representation ensures also the separability of
all $I_k $ (including $H $) in terms of $\zeta _s $. For some of
the best known cases like for the Toda chain, the matrices $L $
and $M $ take values in the normal real form of some simple Lie
algebra $\mathfrak{g} $. This means that $L $ is a linear
combination of the Cartan-Weyl generators with real valued
coefficients.

The complexification procedure renders $p_k $ and $q_k $ complex
and as a result both $L $ and $M $ become complex-valued:
\[
L\to L^\bbbc = L_0 + i L_1, \quad M\to M^\bbbc = M_0 + i M_1,
\quad L, M \in \mathfrak{g}^\bbbc.
\]
Note however, that the complexified Lax equation
\begin{equation}\label{eq:Lax_C}
{dL^\bbbc \over d t} = [L^\bbbc,M^\bbbc],
\end{equation}
can again be written down in terms of purely real Lax matrices
with doubled dimension:
\begin{equation}\label{eq:Lax_2n}
  {d{\bf L}\over d t} = [{\bf L}, {\bf M}], \quad {\bf L} =
\left(
\begin{array}{cc} L_0 & L_1 \\ -L_1 & L_0 \end{array} \right),
\quad {\bf M} = \left( \begin{array}{cc} M_0 & M_1 \\ -M_1 & M_0
\end{array} \right).\nonumber
\end{equation}

This fact we will use below when we discuss generalizations of our
approach.

Let us now consider the Lax representation for the Toda chain
related to the simple Lie algebra $\mathfrak{g} $ with rank $n $:
\begin{eqnarray}\label{eq:L_TC}
L_{\rm TC} = && \sum_{k=1}^{n} p_k H_k +\\&&+ \sum_{k=1}^{n} a_k
(E_{\alpha _k} + E_{-\alpha _k}) + c_0a_0 (E_{\alpha _0} +
E_{-\alpha _0}) , \nonumber\\ M_{\rm TC} = && \sum_{k=1}^{n} a_k
(E_{\alpha _k} - E_{-\alpha _k}) + c_0a_0 (E_{\alpha _0} -
E_{-\alpha _0}) ,\nonumber
\end{eqnarray}
where $\alpha _k $ is the set of simple roots of $\mathfrak{g} $,
$\alpha _0$ is the minimal root of $\mathfrak{g} $,
\begin{equation}\label{eq:a_k}
a_k = {1 \over 2 } \exp ( (\vec{q},\alpha_k )),
\end{equation}
and $c_0 $ was introduced in (\ref{exa:TC}) above. By $E_\alpha $ and
$H_k $ above we denote the Cartan-Weyl generators of $\mathfrak{g}
$; they satisfy the commutation relations:
\begin{eqnarray}\label{eq:CW-cr}
[H_k, E_\alpha ] &=& (\alpha ,e_k)E_\alpha , \nonumber\\ {}
[E_{\alpha }, E_{-\alpha }] &=& H_\alpha ,\\ {}[E_{\alpha },
E_{\beta } ] &=& N_{\alpha ,\beta } E_{\alpha +\beta } ,\nonumber
\end{eqnarray}
where $\alpha $ and $\beta \in \Delta $ are any two roots of
$\mathfrak{g} $ and $N_{\alpha ,\beta }=0 $ if $\alpha +\beta
\not\in \Delta $.

The involution $\mathcal{C} $ induces an involutive automorphism
$\mathcal{C} $ in the algebra $\mathfrak{g} $. Indeed, both $\vec{p} $ and
$\vec{q} $ can be viewed as vectors in the root space $\bbbe^n \simeq
\mathfrak{h}^* $ which is dual to the Cartan subalgebra
$\mathfrak{h}\subset \mathfrak{g}$. Besides, $\mathcal{C} $ must
preserve the Toda chain Hamiltonian:
\begin{equation}\label{eq:H_TC}
H_{\rm TC} = {1 \over 2} \tr L^2_{\rm TC} = {1 \over 2 }
\sum_{k=1}^{n} p_k^2 + \sum_{k=1}^{n} a_k^2 + c_0^2 a_0^2.
\end{equation}
This means that $\mathcal{C} $ in the conformal (resp.\ affine)
case must preserve the system of simple (resp.\ admissible) roots
of $\mathfrak{g} $. This allows one to associate with the
involutive automorphism $\mathcal{C} $ an involutive automorphism
$\mathcal{C}^\# $ of the algebra $\mathfrak{g} $. Obviously, if
$\mathcal{C} $ is defined on $\mathcal{M} $ by
\begin{equation}\label{eq:C_aut}
\mathcal{C}(p_k) = \sum_{s=1}^{n} c_{ks}p_s, \qquad
\mathcal{C}(q_k) = \sum_{s=1}^{n} c_{ks}q_s,
\end{equation}
where $c_{ks} $ are such that $\sum_{s=1}^{n} c_{ks} c_{sm} =\delta _{km}
 $ then the action of $\mathcal{C}^\# $ on $\mathfrak{h} $ can be
determined by duality as:
\begin{equation}\label{eq:C_h}
\mathcal{C}^\# (H_k) \equiv H_{\mathcal{C}(e_k)} =
\sum_{s=1}^{n}c_{ks} H_s.
\end{equation}
On the root vectors $E_\alpha $ the automorphism $\mathcal{C}^\# $
acts as follows:
\begin{equation}\label{eq:C_E}
\mathcal{C}^\# (E_\alpha ) = n_\alpha E_{\mathcal{C}^\# (E_\alpha
) }, \qquad n_\alpha =\pm 1.
\end{equation}

An well known fact \cite{Lie} is that the subgroup of the automorphism
group preserving $\mathfrak{h} $ is determined by
$\Ad_{\mathfrak{h}}\otimes W_{\mathfrak{g}}\otimes V_{\mathfrak{g}} $,
where $\Ad_{\mathfrak{h}} $ is the subgroup of inner automorphisms by
elements of the Cartan subgroup, $W_{\mathfrak{g}} $ is the Weyl group and
$V_{\mathfrak{g}} $ is the group of outer automorphisms. $V_{\mathfrak{g}}
$ is isomorphic to the symmetry group of the Dynkin diagram of
$\mathfrak{g} $.

Since the involution $\mathcal{C} $ for the conformal Toda chain must
preserve the set of simple roots of $\mathfrak{g} $ it must be related to
a symmetry of the Dynkin diagram; therefore $\mathcal{C}^\# $ must
correspond to an outer automorphism of $\mathfrak{g} $. For the affine
Toda chains $\mathcal{C}^\# $ must be related to a symmetry of the
extended Dynkin diagram which allows for a larger set of choices for
$\mathcal{C}^\# $.  These symmetries have been classified in
\cite{Sasaki}.

\begin{remark}\label{rem:7}
The involution $\mathcal{C}^\# $ dual to $\mathcal{C} $ in eq.
(\ref{eq:ex3.1}) is the outer automorphism of $sl(n) $.
\end{remark}

Analogously we consider the Lax representation (\ref{eq:Lax}) of the
Calogero-Moser models is of the form \cite{Perelomov}:
\begin{eqnarray}\label{eq:CMS_Lax}
L(\vec{p},\vec{q}) = \sum_{j=1}^{n} p_j H_j + i
\sum_{\alpha \in \Delta }g_\alpha x((\vec{q},\alpha))E_\alpha , \\
M(\vec{p},\vec{q}) = \sum_{j=1}^{n} z_j H_j + \sum_{\alpha \in
\Delta }g_\alpha y((\vec{q},\alpha))E_\alpha ,
\end{eqnarray}
The constant $g_\alpha$ depends only on the length of the root $\alpha $.
The three functions $x(q) $, $y(q) $ and $z(q) $ satisfy a set of
functional equations
\begin{eqnarray}\label{eq:F_eq}
y(q)=-x'(q), \qquad  z(\xi ) = {x''(\xi)  \over 2x(\xi) }, \\
x(\xi) x'(\eta) - x(\eta_ x'(\xi) = x(\xi+\eta) [z(\xi) - z(\eta)],
\end{eqnarray}
The Hamiltonian $ H= {1\over 2} \tr L^2$ and the function $v(q) $ are
related to them by:
\[ v(q)= -x(q)x(-q).  \]
The solutions to these equations are given by \cite{CMS,Perelomov,Calo}:
\begin{eqnarray}\label{eq:x_xi}
x(\xi) = \left\{ \begin{array}{llll}\displaystyle {1 \over \xi },
\quad &\quad & \quad &\mbox{I} \\
a\,\coth (a\xi), & a\, \sinh^{-1}(a\xi), & & \mbox{II} \\
a\, \cotan (a\xi), & a\, \sin^{-1}(a\xi), & & \mbox{III}
\\ \displaystyle {a\,\cn (a\xi) \over \sn (a\xi)}, & \displaystyle {a\,
\dn (a\,\xi) \over \sn (a\xi)}, & \displaystyle  {a \over \sn (a\xi)}
& \mbox{IV}  \end{array} \right.
\end{eqnarray}
then the corresponding $v(\xi) $ provide the choices in (\ref{eq:ex2_v}).

The involution $\mathcal{C}^\# $ induces a Cartan involution
$\mathop{C}\limits_{\sim} $ on $\mathfrak{g}^\bbbc $ by
$\mathop{C}\limits_{\sim}(Z) =-C(Z^\dag) $ has all the properties of
Cartan involution. As a result the invariance condition with respect to
$\mathop{C}\limits_{\sim} $ restricts to a real form of $\mathfrak{g} $.
The invariance condition for $L $ has the form:
\begin{equation}\label{eq:**}
\mathop{C}\limits_{\sim}\left( L(C^{\#}(\vec{p}),
C^{\#}(\vec{q} ), g\right) = L(\vec{p}, \vec{q}, -g).
\end{equation}
For the case of the example above with $\mathfrak{g}\simeq sl(n) $ and
$x(q) = -x(-q) $ this can be easily shown by realizing that $C(X) $ can be
written in the form $C(X)= TXT^{-1} $ where $T=\sum_{k=1}^{n}E_{k,\bar{k}}
$. One can check that (\ref{eq:**}) and an analogous relation on
$M(\vec{p},\vec{q},g) $ leave the Lax representation invariant. Compare
this with Mikhailov reduction group \cite{Mikhailov}. In this example we
miss the spectral parameter, but we also have the interaction constant $g
$ and (\ref{eq:**}) involves non-trivial action also on $g $.

The substantial difference as compare to the Toda chain case is that the
Calogero-Moser Hamiltonian is invariant with respect to the Weyl group of
$\mathfrak{g}$ as well as with respect to the group of outer automorphisms
of $\mathfrak{g} $. Therefore any involutive element of
$W_{\mathfrak{g}}\otimes V_{\mathfrak{g}} $ can be used to construct a RHF
of the CMS. Obviously two such RHF's will be equivalent if the
corresponding automorphisms $\mathcal{C}_{1} $ and $\mathcal{C}_2 $ belong
to the same conjugacy class of $W_{\mathfrak{g}} $.

This means that the number of inequivalent choices for the involution
$\mathcal{C}$ and therefore, the number of inequivalent RHF of the
Calogero-Moser systems is much bigger. To classify all of them one has to
consider the equivalence classes of $W(\mathfrak{g})$ and pick up just one
element from each class of second order elements. It would be also natural
to relate to each Satake diagram of $\mathfrak{g}$, or a subalgebra of
$\mathfrak{g}$ a RHF of CMS. A detailed study will be reported
elsewhere.

The Cartan subalgebra $\mathfrak{h} $ and the algebra
$\mathfrak{g} $, just like $\mathcal{M} $, can be  splitted into
direct sums:
\begin{equation}\label{eq:g_pm}
\mathfrak{h} = \mathfrak{h}_+ \oplus \mathfrak{h}_- , \qquad
\mathfrak{g} = \mathfrak{g}_+ \oplus \mathfrak{g}_-,
\end{equation}
compatible with the involution $\mathcal{C}^\# $; i.e.:
\begin{eqnarray*}\label{eq:g_pm_C}
\mathcal{C}^\# (X) = X, \quad \mbox{for any }\; X \in
\mathfrak{h}_+ ,  \;\; \mbox{resp., for any }\; X \in
\mathfrak{g}_+ ,\\ \mathcal{C}^\# (Y) = -Y, \quad \mbox{for any
}\; Y \in \mathfrak{g}_- , \;\; \mbox{resp., for any }\; X \in
\mathfrak{h}_+  ,
\end{eqnarray*}
If we compose $\mathcal{C}^\#  $ with $-{}^\dag $ we obtain the
Cartan involution $\widetilde{\mathcal{C}^\# } $ which selects a
real form of the algebra $\mathfrak{g}_\bbbr $.

\section{Dynamics of the Toda chain and its RHF}\label{sec:CTC}

In this Section on the example of the Toda chains, we briefly discuss
how the two basic steps in constructing the RHF's may affect the dynamical
regimes of the Hamiltonian systems.  These two steps are:

a)~complexification and,

b)~imposing the involution $\tilde{\mathcal{C}} $.

The dynamical regimes of the CTC were studied recently from the
point of view of their applications to the adiabatic $N $-soliton
interactions \cite{GUzuEvDi*97}. It was already known that CTC has
a much richer class of dynamical regimes than the real TC. This
shows that step a) (the complexification) changes drastically the
character of the dynamics of out initial Hamiltonian system.

Indeed, it is well known \cite{OlPer*83,OPRS*87} that the solutions of the
conformal Toda chains are parametrized by the eigenvalues $\zeta _j $ and
the first components $r_j $ of the normalized eigenvectors of the Lax
matrix:
\begin{eqnarray}\label{eq:L_j}
L\vec{w}_j =\zeta _j\vec{w}_j , \qquad r_j=\vec{w}_j^{(1)}, \\
(\vec{w}_j,\vec{w}_k) \equiv \sum_{s=1}^{n}\vec{w}_j^{(s)} \vec{w}_k^{(s)}
= \delta _{jk}.
\end{eqnarray}
For the Toda chain related to the algebra $sl(n) $ these solutions take
the form:
\begin{eqnarray}\label{eq:q_k}
q_1(t) &=& \ln A_1(t), \qquad A_1(t)= \sum_{j=1}^{n} r_j^2 e^{-2\zeta
_jt}, \nonumber\\
q_k(t) &=& \ln {A_{k+1} \over A_k }, \qquad k=2,3,\dots,n, \\
\label{eq:A_k}
A_k(t) &=& \sum_{j_1<\dots <j_k}^{} W^2(j_1,\dots , j_k) r_{j_1}^{2} \dots
r_{j_k}^{2} \times\\& \times&\exp (-2(\zeta _{j_1} + \dots + \zeta
_{j_k})t)\nonumber\\
A_n&=& \prod_{k=1}^{n}r_j^2 W^2(1,2,\dots,n), \qquad A_{n+1}=1,\nonumber
\end{eqnarray}
where $W_{j_1,\dots, j_k} $ is the Wandermonde determinant:
\begin{equation}\label{eq:W_k}
W_{j_1,\dots, j_k} = \det \left( \begin{array}{cccc} 1 & 1 & \dots & 1 \\
\zeta _{j_1} & \zeta _{j_2} & \dots & \zeta _{j_k} \\ \vdots & \vdots &
\ddots & \vdots \\ \zeta _{j_1}^{k-1} & \zeta _{j_2}^{k-1} & \dots & \zeta
_{j_k}^{k-1} \end{array} \right).
\end{equation}

It is well known that for the real Toda chain:

1) all $\zeta _j $ are real and pair-wise different;

2) all $r_j $ are also real \cite{Moser}.

This ensures that $A_k(t) $ are positive for all $t $ and therefore the
solutions $q_k(t) $ are regular for all $t $. In addition one can evaluate
the asymptotic behavior of $q_k(t) $ for large $|t| $. If we assume that
the eigenvalues $\zeta _k $ satisfy the sorting condition:
\begin{equation}\label{eq:sc}
\zeta _1>\zeta _2>\dots > \zeta _n,
\end{equation}
then we get the result:
\begin{eqnarray}\label{eq:qk-as}
\lim_{t\to \pm\infty } (q_k(t) - v_{k}^{\pm}t ) =\beta _{k}^{\pm}, \\
v_k^- =-2\zeta _k, \qquad v_k^+ = -2\zeta _{n+1-k};
\end{eqnarray}
for the constants $ \beta _{k}^{\pm}$ one can derive explicit expressions
in terms of $\zeta _j $.

If we interpret $q_k(t) $ as the trajectory of the $k $-th
particle then $2\zeta _j $ characterize their asymptotic velocity.
The property 1) above means that the (real) Toda chain allows only
for (non-compact) asymptotically free motion of the particles.

The set of coefficients $\{\zeta _j, \ln r_j\}$ are the
action-angle variables in the generalized sense, introduced in the
beginning of Section 5.  Of course they are convenient for solving
the TC model because:  i) they satisfy canonical Poisson brackets
and ii) the TC Hamiltonian takes the simple form:
\begin{equation}\label{eq:HTC_zeta}
H_{\rm TC} = \sum_{j=1}^{n} 2\zeta _j^2.
\end{equation}
Therefore the equations of motion for $\{\zeta _j, \ln r_j^2\}$ take the
form:
\begin{equation}\label{eq:eq_mot}
{d\zeta _j  \over dt } =0, \qquad {d\ln r_j^2  \over dt } = -2\zeta _j,
\qquad j=1,\dots, n.
\end{equation}

The situation changes substantially after the complexification, see
\cite{GEI}.  In fact the same formulae (\ref{eq:q_k}), (\ref{eq:A_k})
provide the solution also for the complex case. But now we have neither of
the properties 1) or 2).  The eigenvalues $\zeta _j $, as well as $r_j $
become complex:
\begin{equation}\label{eq:z_c}
\zeta _j = \kappa _j + i\eta_j, \qquad \rho _j+i\phi _j = \ln
r_je^{q_1(0)}, \qquad j=1,\dots, n,
\end{equation}
where
\[ e^{-nq_1(0)} = A_n = \prod_{j=1}^{n}r_j^2 W^2(1,\dots,n).\]
This substantially modifies the properties of the solutions.

For CTC one can derive the analogs of the relations (\ref{eq:qk-as}):
\begin{eqnarray}\label{eq:qkc-as}
\lim_{t\to \pm\infty } (q_k(t) - v_{k}^{\pm}t ) =\beta _{k}^{\pm}, \\
v_k^- =-2\kappa _k, \qquad v_k^+ = -2\kappa _{n+1-k};
\end{eqnarray}
which shows that now it is the set of real parts $\{\kappa _j\} $ of the
eigenvalues $\zeta _j $ which determine the asymptotic velocities.
Even if we assume\footnote{In principle we may have $\zeta _j=\zeta _k $
for $k\neq j $ which leads to degenerate solutions; we will not consider
such cases below, see e.g. \cite{GEI}.} that $\zeta _j\neq \zeta _k $ for
$j\neq k $ then we still may have $\kappa _j = \kappa _k $; note that
equal asymptotic velocities allow for the possibility of bound states.
Skipping the details (see \cite{GUzuEvDi*97,GEI}) we just list the
different types of asymptotic regimes for the CTC which are determined by
the so-called sorting condition:
\begin{equation}\label{eq:scc}
\kappa _1\geq \kappa _2\geq \dots \geq \kappa _n.
\end{equation}
This sorting condition differs from (\ref{eq:sc}) for the real TC
in that it allows equalities. Therefore the CTC allows several
non-degenerate regimes:

A) Asymptotically free regime: $\kappa _j \neq \kappa _k $ for $j\neq k $;

B) $n $-particle bound state for $\kappa _1 = \kappa _2 = \dots = \kappa
_n $;

C) A number of mixed regimes when we have two or more groups of equal
$\kappa _j $'s;

Another difference with respect to the real TC case is that $A_k(t) $ are
no more positive definite. For some choices of the $\zeta _j $ and $r_j $
they may vanish even for finite values of $t=t_0 $ which means that
$q_k(t) $ may develop singularity for $t\to t_0 $.

Let us now analyze the effect of the involutions $\mathcal{C} $
and $\tilde{\mathcal{C}}\equiv \mathcal{C}\circ \ast $ on the spectral data
$\{\zeta _k,\ln r_k\} $ of $L $.  It will allow us to describe the
dynamical regimes of the RHF of CTC. Eq. (\ref{eq:ex3.1}) leads to:
\begin{eqnarray}\label{eq:Ct_sln}
\tilde{\mathcal{C}} (q_k) = - q_{n+1-k}^*, \qquad  \tilde{\mathcal{C}}
(p_k) = - p_{n+1-k}^*,
\end{eqnarray}
and consequently:
\begin{equation}\label{eq:a_b_inv}
\tilde{\mathcal{C}} (b_k) = - b_{n+1-k}^*, \qquad  \tilde{\mathcal{C}}
(a_k) = a_{n-k}^*.
\end{equation}
These constraints mean that the Lax matrix $L $ satisfies:
\begin{eqnarray}\label{eq:5.5}
S_0 L(\tilde{\mathcal{C}} (p_k),\tilde{\mathcal{C}} (q_k)) S_0^{-1} = -
L(p_k,q_k)^\dag, \\ S_0= \left( \begin{array}{ccccc} 0 & 0 & \dots & 0 & 1
\\ 0 & 0 & \dots & -1 & 0 \\ \vdots & \vdots & \ddots& \vdots & \vdots \\
0 & (-1)^{n-1} & \dots & 0 & 0 \\ (-1)^{n} & 0 & \dots & 0 & 0 \end{array}
\right).
\end{eqnarray}
This means that the eigenvalues and the matrix of eigenvectors of
$L $ satisfy:
\begin{eqnarray}\label{eq:5.6}
S_0 Z S_0^{-1} &=& - Z^*, \qquad Z = \diag (\zeta _1,\dots, \zeta
_n)\;
\mbox{ or} \nonumber\\
\tilde{\mathcal{C}} (\zeta _k) &=& - \zeta _{n+1-k}^*.
\end{eqnarray}
Technically it is a bit more difficult to derive the action of
$\tilde{\mathcal{C}} (r _k) $. Indeed, this must ensure that the functions
$A_k $ satisfy
\begin{equation}\label{eq:A_k2}
\tilde{\mathcal{C}} (A_k) = A_{n-k}^*,
\end{equation}
for all values of $t $. After some calculations we get that the relations
\begin{eqnarray}\label{eq:Ct_rk}
\rho _{n+1-k} + i\phi _{n+1-k} &=& -\rho _k +i\phi _k + \ln w_k ,\\
w_k&=& {W(1,\dots,k-1,k+1,\dots,n) \over W(1,2,\dots,n) }\nonumber
\end{eqnarray}
ensure that $B_{n-k}^* = B_k $ where
\begin{equation}\label{eq:Ct_rk'}
B_k (t)= A_k(t)e^{-kq_1(0)}.
\end{equation}

Let us illustrate this by first putting $n=4 $ and choosing
\begin{eqnarray}\label{eq:*0}
\zeta _1 &=& -\zeta _4^* =\kappa _1 + i \eta_1,\qquad
\zeta _2=-\zeta _3^* =\kappa _1 - i \eta_1, \nonumber \\
\rho _1&=&\rho _0 + i \phi _0, \qquad
\rho _3=-\rho _0 - i (\phi _0+\alpha + {\pi  \over 2 }), \\
\rho _2&=&\rho _0 - i \phi _0, \qquad
\rho_4=-\rho _0 +i (\phi _0+\alpha + {\pi  \over 2 }). \nonumber
\end{eqnarray}

With these notations we can write down the solution for the $sl(4) $-CTC
in the form:
\begin{eqnarray}\label{eq:*1}
q_1(t)&=&\ln B_1(t), \qquad q_2(t) = \ln {B_2(t)  \over B_1(t) },\\
q_3(t)&=&\ln {B_3(t)  \over B_2(t) }=-q_2^*(t), \qquad q_4(t)=-\ln B_3(t)
=-q_1^*(t), \nonumber
\end{eqnarray}
where
\begin{eqnarray}\label{eq:*2}
B_1(t)&=& {1  \over 2^5\kappa _1\eta_1\sqrt{\kappa _1^2+\eta_1^2} } \left[
e^{-2\kappa _1t + 2\rho _0} \cos(2\eta_1 t - 2\phi _0) \right. \nonumber\\
&-& \left.  e^{2\kappa _1t - 2\rho _0} \cos(2\eta_1 t - 2\phi _0-2\alpha
)\right] \\
B_2(t)&=& {-1  \over 2^7\kappa _1^2\eta_1^2(\kappa _1^2+\eta_1^2) }
\left[ \eta _1^2 \cosh(4\eta_1 t - 4\phi _0) \right. \nonumber\\
&+&\left. \kappa _1^2 \cos(4\eta_1 t - 4\phi _0-2\alpha )+ \kappa _1^2
+\eta_1^2\right] , \nonumber\\
B_3(t) &=& B_1^*, \qquad \alpha _1 = \arg \zeta _1 = \arctan {\eta_1
\over \kappa _1}.  \nonumber
\end{eqnarray}

The choice of $\zeta _1=\zeta _2^* $ in (\ref{eq:*0}) combined with the
involution $\zeta _1=-\zeta _4^* $, $\zeta _2=-\zeta _3^* $ ensures that $
q_1(t) $ and $q_2(t) $ will have equal asymptotic velocities and will form
a bound state. Indeed, from (\ref{eq:*1}) and (\ref{eq:*2}) we get:
\begin{eqnarray}\label{eq:*3}
&& \lim_{t\to\pm\infty }\left( {1  \over 2 }(q_1(t)+q_2(t)) \mp 2\kappa _1t
\right) = {i\pi  \over 2 } \mp 2\rho _0 \nonumber\\
&& \qquad - {1  \over 2 } \ln  2^8\kappa _1^2 (\kappa _1^2 + \eta_1^2) +
\mathcal{O}\left( e^{-4\kappa _1|t|}\right),
\end{eqnarray}
which means that the center of mass of the particles $q_1(t) $  and
$q_2(t) $ for $t\to\pm\infty  $ undergoes non-compact asymptotically free
motion with asymptotic velocities $\pm 2\kappa _1 $. At the same time the
relative asymptotic motion is compact:
\begin{eqnarray}\label{eq:*4}
q_1(t)&-&q_2(t) \mathop{\longrightarrow}\limits_{t\to\pm\infty } i\pi -
\ln 4\eta_1^2 + \ln \cos^2 (\Phi ^\pm(t))\\
&& \qquad + \mathcal{O}\left( e^{-2\kappa _1|t|}\right),\nonumber\\
\Phi ^+(t)&=& 2\eta_1t - 2\phi _0 -2\alpha _1, \qquad
\Phi ^-(t)= 2\eta_1t - 2\phi _0 . \nonumber
\end{eqnarray}

The energy of these RHF of $sl(4) $-CTC is given by:
\begin{equation}\label{eq:*5}
H_{sl(4)} = 8(\kappa _1^2 -\eta_1^2).
\end{equation}

Due to the symmetry the `particles' $q_3(t) $ and $q_4(t) $ also form a
bound state; the corresponding formulae easily follow from (\ref{eq:*3})
and (\ref{eq:*4}) and $q_4(t)=-q_1^*(t) $ and $q_3(t)=-q_2^*(t)  $.

This example demonstrates several peculiarities of the RHF dynamics. The
first one is that configurations in which all $n $ particles form a bound
state is impossible. Indeed, such bound state may take place if  $\kappa
_j=0 $ for all $j=1,\dots,n $. But it is easy to check that the sorting
condition (\ref{eq:scc}) and the symmetry of the eigenvalues (\ref{eq:*0})
lead to degeneracy of the spectrum of $L $ for $\kappa _j=0 $.

The RHF dynamics allows either asymptotically free regimes or mixed
regimes with special symmetry, namely we can have only even number of
bound states which move with opposite asymptotic velocities.

The above considerations on the properties of the dynamical regimes are
not specific for the Toda chain.  Similar conclusions can be derived  also
for the RHF of the Calogero-Moser systems. Here we have richer variety of
RHF's due to the fact that the CMS Hamiltonian is invariant with respect
to the whole Weyl group of $\mathfrak{g} $. More detailed investigation of
the CMS dynamical regimes will be done elsewhere.

\section{Discussion}\label{sec:Dis}

\begin{enumerate}
\item The method goes also for non-integrable models, though we
treated mainly integrable ones.

Other related aspects of our method concern more complicated
objects such as dynamical classical $r $-matrices or quantum $R
$-matrices due to their algebraic structures will also satisfy
relations of the form (\ref{eq:**}).

\item Our list of examples can be substantially extended with
Ruijsenaars-Sneider models and their generalizations being among
the obvious candidates.

\item The present approach can naturally be generalized also for
infinite-dimensional integrable models such as the $1+1
$-dimensional Toda field theories. This will allow one to
construct new classes of real Hamiltonian forms of Toda field
theories extending the results of \cite{Evans}.

\item The method can be used repeatedly: Since the complexified
model can be viewed as real Hamiltonian system with $2n $ degrees
of freedom, then we can start and complexify it getting a
Hamiltonian system with $4n $ degrees of freedom producing a real
form dynamics with $2n $ degrees of freedom.

If the initial system allows Lax representation that depends analytically
on the dynamical parameters:
\begin{equation}\label{eq:L0-M0}
{dL_0 \over d t} = [L_0,M_0].
\end{equation}
After the complexification we get complex-valued $L=L_0+iL_1 $ and
$M=M_0+iM_1 $ which provide the Lax representation for the complexified
system. But the complexified system can be viewed as real Hamiltonian
system with $2n $ degrees of freedom. It allows Lax representation with
real-valued $L $ and $M $ of the form (\ref{eq:Lax_2n}).

\item Solutions of the initial system that depend analytically on
the initial parameters `survive' the complexification procedure.
As we see from the CTC case discussed above, solutions that have
been regular for all time after such procedure may develop
singularities for finite $t$. Another substantial difference is in
the asymptotic dynamical regimes.

Projecting onto the RHF some of these singularities remain. Their
asymptotic dynamical regimes need additional studies.
\end{enumerate}

\section{Acknowledgements}

One of us (VSG) acknowledges financial support in part from Gruppo
collegato di INFN at Salerno, Italy and proj\-ect PRIN 2000
(contract 323/2002). The work of GM and GV was partially supported
by PRIN SINTESI. One of us (VSG) thanks Professor F. Calogero for
the hospitality at Rome University and the nice discussions
concerning this paper. One of us (AK) is grateful to the Department of
Physics and to  for the hospitality of the University of Salerno and to
INFN group at Salerno for their warm hospitality.

\end{document}